\documentclass[10pt]{article}

\usepackage{amsfonts}
\usepackage[dvips]{graphicx}
\usepackage{color}

 \newtheorem{remark}{Remark}[section]
 
\newcommand\eq[1] {(\ref{#1})}
\begin{document}

\title{Optimal anisotropic three-phase conducting composites. Plane problem}

\author{Andrej Cherkaev and  and Yuan Zhang \\
~\\Department of Mathematics, University of Utah, \\Salt Lake City, Utah, USA, }
\maketitle
\abstract{
The paper establishes tight lower bound for effective conductivity tensor $K_*$ of two-dimensional three-phase conducting anisotropic composites and defines optimal microstructures. It is assumed that three materials are mixed with fixed volume fractions and that the conductivity of one of the materials is infinite. The bound expands the Hashin-Shtrikman and Translation bounds to multiphase structures, it is derived  using  a combination of Translation method and additional inequalities on the fields in the materials; similar technique was used by Nesi (1995) and Cherkaev (2009)  for isotropic multiphase composites. This paper expands the bounds to the anisotropic composites with effective conductivity tensor $K_*$. The lower bound of conductivity (G-closure) is a piece-wise analytic function of eigenvalues of $K_*$, that depends only on conductivities of components and their volume fractions. Also, we find optimal microstructures that realize the bounds, developing the technique suggested earlier by Albin et al. (2007) and Cherkaev (2009). The optimal microstructures are laminates of some rank for all regions. The found structures match the bounds in all but one region of parameters; we discuss the reason for the gap and numerically estimate it.
}

~


\section{Introduction}
\paragraph{The problem}
The paper investigates the lower bound for effective conductivity and optimal micro-geometries of three-material composites (plane problem). We  assume that two mixing isotropic materials have finite conductivities $k_1$ and $k_2$, ($0<k_1<k_2$) and the third one is a superconductor $k_3=\infty$, the volume fractions $m_1 \geq 0, m_2 \geq 0 $ and $ m_3=1-m_1-m_2 \geq 0$ of the materials are fixed. The conductivity of a composite is characterized by an anisotropic effective conductivity tensor $K_*$ that depends on the properties of mixed materials and their volume fractions, as well as on microstructures. We describe the bounds of {\em G-closure} (Lurie and Cherkaev, 1981) -- the set of all effective properties of composites with arbitrary microstructure. The G-closure boundary depends only on $k_1, k_2, m_1, m_2$. Optimal  microstructures {\em realize the bound} if their effective conductivity lies at the G-closure boundary.  

We find the bound solving a variational problem of  minimization of $K_*$ with respect to microstructures (Section \ref{s:problem}).
Namely, we apply two orthogonal external fields of different magnitudes to a periodic composite and minimize the sum $J$ of the corresponding energies of the composite, varying the microstructure occupying  the periodicity cell $\Omega$. The computed value of $J$ allows for computation of the   add outer bound of G-closure, as discussed in Sections \ref{symm} and \ref{s:g-cl}.  
The matching microstructures (minimizing sequences) are found by a different technique that was introduces in (Albin et al, 2007a, Cherkaev, 2009) and is described in Sections \ref{s:th4} and \ref{s:struct};  by assumption, optimal structures are laminates of some rank. The effective properties of the structures form the inner bound of G-closure. When the outer and inner bounds coincide, they are exact and the G-closure is determined. We show that our bounds are exact in all domains of parameters but one. In the last domain, we estimate the gap between the outer and inner bounds. 
\begin{remark}
 The complementary upper bound can be  established by a solution of a dual problem, in which conductivity $k_i$ are replaced by resistivity $\rho_i=1/k_i$. In the considered problem, one of the component is a superconductor  ($k_3=\infty$) which makes the dual bound trivial - the effective resistivity can be arbitrary large, or $K_*^{-1} \geq 0$. The obtained results allows for the upper bound determination for the G-closure of materials with conductivities $k_1=0 < k_2< k_3 < \infty$.
 \end{remark}

\paragraph{Bounds}

The problem of exact bounds has a long history. It started with the bounds by Voigt (1928) and Reuss (1929), called also Wiener bounds or the arithmetic and harmonic mean bounds. The bounds are valid for all microstructures and become in a sense exact for laminates: One of the eigenvalues of $K_*$ of a laminate is equal to the harmonic mean of the mixed materials' conductivities, and the other one - to the arithmetic mean of them.  The pioneering paper by Hashin and Shtrikman (1963) found the bounds and the matching structures for optimal isotropic two-component composites, and suggested bounds for multicomponent ones. The exact bounds and optimal structures of  {\em anisotropic} two-material composites were found in earlier papers  (Lurie and Cherkaev,1982, 1986, Kohn and Strang, 1983, 1986, Tartar, 1985) using a version of the translation method (see its description in books (Cherkaev, 2000, Allaire, 2001, Milton, 2002, Dacorogna, 2008)). The method is equivalent  to building the polyconvex envelope of a multiwell Lagrangian, as it was shown by Kohn and Strang (1983, 1986); the wells are the energy of the materials plus their cost (here, ``cost" is the dual variable to the volume fraction of material in the composite).   The theory of bounds for the two-material composite is now well developed and applied to elastic, viscoelastic, and other linear materials, see for example the books (Lurie, 1993,  Cherkaev, 2000, Allaire, 2001, Milton, 2002, Dacorogna, 2008). 

Bounds for multicomponent composites turn out to be much more difficult. Milton (1981) showed that the Hashin-Shtrikman bound is not exact everywhere (it tends to an incorrect limit when $m_1\to 0$), but is exact when $m_1$ is larger than a threshold, $m_1\geq g_m$.
Milton and Kohn (1988) suggested an extension of the translation method to anisotropic multimaterial composites, computed the anisotropic bounds for multicomponent composites and the optimal structures. 
Nesi (1995) suggested a new tighter bound for {\em isotropic} multicomponent structures, and Cherkaev (2009) further improved it and found optimal structures. The method is based on the procedure suggested by Nesi (1995) that combines the translation method and additional inequality constraints (Alessandrini and Nesi, 2001).  These two latest bounds coincide in the case $k_3=\infty$ that is considered here.  

This paper extends these bounds to {\em anisotropic} composites. As in the early paper by Kohn and Strang (1983), we investigate the case when one of the phase has infinite conductivity, which significantly simplifies the calculation. The method is based on constructing a lower bound  for the composite energy accounting for  Alessandrini and Nesi (2001)  constraints. Because of the constraints, the translated energies-wells can become nonconvex but are still bounded from below, an improved bound corresponds to this case. The method is described in Section \ref{locpoly}, the results are summarized in Section \ref{s:results}. The energy bound turns out to be a multi-faced piece-wise analytic function of the problem's parameters. Like the translation bound,  it depends only on conductivities of the materials, their volume fractions, and the anisotropy of a homogeneous external loading.  The  energy bounds and related bounds for the G-closure are derived in Section \ref{s:bounds}.

\paragraph{Optimal structures}

In the paper, we prove that multiscale laminates realize the G-closure bound. Similar structures - laminates of second rank - realize the G-closure bound for the two-material case (Lurie and Cherkaev, 1982, 1986); three-material bound is achievable by more complex structures of the same kind.   Optimal structures depend on the degree of anisotropy of the external loading.
\begin{remark}
Generally, optimal structures are not necessary laminates: For example, Hashin and Shtrikman (1963) first suggested ``coated spheres" geometry,  Milton (1981)  and parallel coated spheres and later suggested (Milton, 2002) a method of transformation of optimal shapes, Lurie and Cherkaev (1986) suggested multilayer coated circles, Vigdergauz (1989),  Grabovsky and Kohn (1995) and recently  Lui (2008) suggested special convex oval-shaped inclusions, Gibiansky and Sigmund (2000) suggested "bulk blocks", Albin and Cherkaev (2006) proved the optimality of ``haired spheres", and recent paper by Benveniste and Milton (2010) investigated "coated ellipsoids". All these structures admit separation of variables when  effective properties are computed. It is not clear yet if the laminate structure approximates any other optimal structure, see for example (Pedregal,1997, Briane and Nesi, 2004, Albin et al., 2007b). We show, however, that proper laminates are optimal for the considered problem. 
\end{remark}
The topology of two-material optimal structures is simple and intuitively clear: for isotropic or moderately anisotropic  loading, the less conducting  material  $k_1$ ``wraps" the more conducting one $k_2$ ($k_2>k_1$), so that $k_2$ forms an nucleus and $k_1$ forms  a core. If the anisotropy of the loading exceeds a threshold, the optimal structures degenerate into  simple laminates. 

The multimaterial structures are more diverse and nonunique and require new ideas for constructing.  
Milton (1981), Lurie and Cherkaev (1985) and later Barbarosie (2001) described two types of isotropic structures that realize the multicomponent bound for sufficiently large volume fractions $m_1 \geq g_{m} $ of the  weaker conductor $ k_1 < k_2< \ldots $  Later, Gibiansky and Sigmund (2000) expand the domain of applicability of Hashin-Shtrikman bounds to $m_1 \geq g_{gs}$ where the threshold $  g_{gs}$ is smaller than the one previously known $  g_{gs}< g_m$. They demonstrated new isotropic non-laminate microstructures (bulk structures) that realize this bound. Liu (2008) found another structures an optimal conductivity. 
 Albin et al. (2007) extended the results of Gibiansky and Sigmund, (2000) finding anisotropic laminates that realize translation bounds for both isotropic and anisotropic structures in a range of parameters
$$m_1 \geq g_{acn},  \quad \mbox{and }  \frac{|k_{*2}-k_{*1}|}{k_{*1}+k_{*2}} \leq \hat{g}_{acn}$$ 
 where $ k_{*1}$ and $k_{*2}$ are eigenvalues of $K_*$.   These inequalities restrict the range of volume fractions  and degree of anisotropy of a composite that correspond to translation bounds.  For isotropic composites ($k_{*1}=k_{*2}$), the range of applicability of the founded laminates coincides with the one of bulk structures founded by Gibiansky and Sigmund (2000).  
 Structures that realize the isotropic bound for the whole range of volume fractions were found in (Cherkaev, 2009). 

  In Section \ref{s:struct} we extent this result to anisotropic composites, finding new optimal structures that realize our new bounds. More exactly, we show that optimal laminates realize the bounds in all but one region.  The topology of optimal structures depends on volume fractions of the mixing elements and loading anisotropy level. The structure adjusts  itself to meet the {\em sufficient} optimality conditions that are found during derivation of the bounds. All the optimal microstructures are found by the same procedure suggested in (Albin et al, 2007)) and based on 
(i) the energy bounds and sufficient optimality conditions for gradient fields inside each material, and 
(ii) the lamination technique that allows for satisfaction of these conditions.
In all cases but one, the found laminate achieve the bounds, they are not unique.

In the remaining case, the lower bound for G-closure is definitely not exact, hence the mentioned technique for building the structures is not applicable. In that case, we guess the best structures (that correspond to the upper bound of G-closure) basing on asymptotic behavior of optimal structures in neighboring regions and then numerically compute the gap between the structures and bound that is between the upper and lower bounds for G-closure). The gap is very small, see Section \ref{s:structE}, which shows that the suggested laminates  (upper bound) and the lower bound accurately approximate G-closure.

\section{The problem}\label{s:problem}

\subsection{Equations and notations}

Consider a periodic composite formed by three materials.
The materials $k_i$ occupy plane domains $\Omega_i, ~i=1,2,3 \subset R_2$ that form a unit periodicity cell  $\Omega$ 
$$
\bigcup_{i=1,2,3} \Omega_i= \Omega, \quad  \Omega=\{ (x_1, x_2):~ 0 \leq x_1 < 1, ~  0 \leq x_2 < 1\} .
$$ 
The areas $m_i =\| \Omega_i \| $ of $\Omega_i$ are fixed:
 $ m_1+m_2+m_3=1, ~m_i \geq 0$, and there are no other constraints on $\Omega_i$.

The conductivity of the composite is described by the system of equations 
that expresses the curlfree nature of the field $e$, the divergencefree nature of the currents $j$, and the constitutive relation (Ohm's law) that joins the field and current
\begin{eqnarray}
 \nabla \times e=0, \quad \mbox{or } e=-\nabla u, ~~ \quad  \nabla \cdot j=0 \quad  \mbox{in } \Omega,
\quad  j=k_i e ~~  \mbox{in } \Omega_i  \label{e-diff}
  \end{eqnarray}
where $u\in H_1(\Omega)$ is a scalar potential. Potential $u$ and the normal current are continuous at the boundaries $ \partial_{ik} $ between $\Omega_i$ and $\Omega_k$, the conditions hold
\begin{equation}
\tau \cdot (e_i-e_k)=0, \quad \nu \cdot (j_i-j_k)=0 \quad \mbox{at } \partial_{ik} 
\label{bound-cond}
\end{equation}
where $\tau$ and $\nu$ are the tangent and normal vectors to $\partial_{ik} $.

In order to determine effective properties of the composite, we subject it to two homogeneous orthogonal {\em external  loadings}  $e_{0a}$ and $e_{0b}$,
$$
\lim_{|x|\to \infty}e_a(x)= e_{0a}=\pmatrix{1 \cr 0} \quad \mbox{and } \lim_{|x|\to \infty}e_b(x)=e_{0b}=\pmatrix{0 \cr r}
$$
where $0\leq r \leq 1$ and calculate the sum of energies. Thus, we deal with 
a couple of conductivity equations (\ref{e-diff}) (denoted with subindexes $a$ and $b$) that differ only in boundary conditions.
This couple can be conveniently viewed as a problem for a vector potential $U=(u_a, u_b)^T$ and $2 \times 2$ matrices $E=(e_a | e_b)$ and $J=(j_a | j_b)$ in a cell $\Omega$:
\begin{eqnarray}
  \nabla \cdot J=0, ~~   E=-\nabla U \mbox{ in } \Omega, \quad J= k_i\, E \mbox{ in } \Omega_i   \quad \lim_{|x|\to \infty}  E(x) = E_0 .
\label{vars}
\\
E(x) \mbox{ is periodic in }\Omega, \quad  \int_{\Omega} E(x) = E_0, \quad  
   E_0= \pmatrix{1 & 0 \cr 0 & r}
\label{E0-def}  \label{aver}
\end{eqnarray}

\paragraph{Energy}

In each material $k_i$,  energy $W_i={W_i}_a + {W_i}_b$  of the coupled conductivity equation is defined as a sum of two energies ${W_i}_a$ and ${W_i}_b$, caused by the  loadings $e_{0a}$ and $e_{0b}$, respectively,
\begin{eqnarray*}
 W_i=  {W_i}_a+{W_i}_b= \frac{1}{2} k_i \mbox{Tr} (E\,E^T)
\end{eqnarray*}
where $ \mbox{Tr}$ denotes the trace. The energy density of an anisotropic material with conductivity tensor $K$ has a similar form 
\begin{equation}
 W(K, E)=   \frac{1}{2}  \mbox{Tr}(K \,E\,E^T)
\label{11}
\end{equation}
Here, we assume that material 3 is a superconductor, $k_3=\infty$, as in  Kohn and Strang (1983). In $ \Omega_3 $, field $E$ is zero and the current is not defined. The energy of the superconductor is 
\begin{equation} 
W_3(E)=\left\{\begin{array}{ll}
0 & \mbox{if }E=0 \\
+\infty & \mbox{if }E \neq 0 
\end{array} \right. .
\label{en3}
\end{equation}

The energy $W_0(E_0)$ of the whole periodicity cell has the form
\begin{equation}
W_0(E_0)=\frac{1}{2}  \mbox{Tr} (K_*E_0\,E_0^T)=\inf_{E {\rm \,as \,in\,} ( \ref{vars}), (\ref{aver}) } \sum_{i=1}^2\int_{ \Omega_i}W_{i}(E) dx,
\label{en-def}
\end{equation}
it defines the effective tensor $K_*$ and depends on parameters $k_i, m_i, r$ and on subdivision $\Omega_i$.

\paragraph{Basis}
It is convenient to decompose the $2\times 2$ matrix $E$ using an orthonormal basis 
 \begin{eqnarray}
{\bf a_1} &=&  \frac{1}{\sqrt{2}} \pmatrix{1 & 0 \cr 0 & 1} , ~~ {\bf a_2}=  \frac{1}{\sqrt{2}} \nonumber
\pmatrix{1 & 0\cr 0& -1}, \nonumber \\
 {\bf a_3} &=&  \frac{1}{\sqrt{2}}  \pmatrix{0 & 1 \cr -1 & 0} 
  ~~ 
  {\bf a_4} =  \frac{1}{\sqrt{2}}  \pmatrix{0 & 1 \cr 1 & 0} \label{basis}
 \end{eqnarray}
The decomposition takes the form
 \begin{equation}
  E=  \frac{1}{\sqrt{2}}s_1(E) {\bf a_1} + \frac{1}{\sqrt{2}}d_1(E) {\bf a_2} + \frac{1}{\sqrt{2}}d_2(E) {\bf a_3}  + \frac{1}{\sqrt{2}}s_2(E) {\bf a_4}
  \label{decom1}
 \end{equation}
 where
 \begin{eqnarray}
& & s_1(E)= \frac{1}{\sqrt{2}}(E_{11}+ E_{22}), \quad 
d_1(E)=\frac{1}{\sqrt{2}}(E_{11}- E_{22})
\nonumber \\
& & s_2(E)= \frac{1}{\sqrt{2}}(E_{12}- E_{21}),\quad
d_2(E)= \frac{1}{\sqrt{2}}(E_{12}+ E_{21}).
\label{x2}
\end{eqnarray}
 The energy has a form
\begin{equation}
W(k,E)= \frac{1}{2} {k} Tr(E \,E^T)= \frac{1}{2}k (s_1^2 + s_2^2+ d_1^2+ d_2^2).
\label{en1,2}
\end{equation}
Notice the representation for determinant
\begin{equation}
 \det(E) =\frac{1}{2} (s_1^2+ s_2^2 -d_1^2-d_2^2). 
 \label{det-repr}
 \end{equation}

In the next consideration, we assume that the  average field is fixed: $(s_1, s_2, d_1, d_2) \in {\cal E}_{av}$, where
\begin{eqnarray}
{\cal E}_{av}= 
& &\left\{ E: \int_\Omega s_1(x) \,dx=S_{01}=\frac{(1+r)}{\sqrt{2}}, \quad \int_\Omega s_2(x) \,dx=S_{02}=0,  \right.
\nonumber \\
& &\left. \int_\Omega d_1(x) \,dx=D_{01}=\frac{(1-r)}{\sqrt{2}} ,~~\int_\Omega d_2(x) \,dx=D_{02}=0 \right\} 
\label{e0} 
\end{eqnarray}

\subsection{Bounds}\label{symm}
\paragraph{Energy Bounds}
We find a lower geometrically independent bound for $W_0$ by arbitrarily varying subdomains $\Omega_1$ and $\Omega_2$ but preserving their areas -- fractions $m_i$ of the materials in the composite. 
The bound depends on ratio $r$ of the external loadings. Without lose of generality, we assume that the largest loading equals one, $0 \leq r \leq 1$.
The case  $r=0$ corresponds to only one loading, and $r=1$ corresponds to two loadings of equal magnitude.
The bound has the form, see \eq{en-def}
\begin{equation}
B(E_0 k_i,m_i )
=\inf_{\Omega_i: |\Omega_i|=m_i } W_0(E_0).
\end{equation}
where $E_0(r)$ is given by \eq{e0}.
Below, we write $B=B(r)$, assuming that other parameters are fixed. Notice that sum of the energies $B$ stays constant if the order of external loadings is reversed, or labeling of the axes is reversed, or $r$ changes its sign.

Algebraically, the bound $B$ can be expressed through a tensor $K_*(E_0)$ of effective properties of the optimal composite with eigenvalues $k_{*1}$ and $k_{*2}$ and eigenvectors directed along $OX_1$ and $OX_2$ axis, respectively.

\begin{equation}
2 B(E_0, k_i,m_i, r)
=k_{*1}+ k_{*2} r^2
\end{equation}

\subsection{G-closure boundary} \label{s:g-cl}
The bound for the energy implies the bound for the effective properties tensor $K_*$. To derive this bound, we use representation \eq{11}
\begin{equation} 
2W_{comp}(r)= \left(k_{*1} + k_{*2} r^2\right)\geq 2B(r)
\label{comp-energy}
\end{equation}
as a function of $r$.
If the  eigenvalues $k_{*1} , k_{*2}$ lie on the boundary of G-closure, the inequality for the bound becomes an equality, 
$W_{comp}(r)-B(r)=0.
$ 

Each value of $r$ corresponds to an optimal pair $\left(k_{*1}(r) , k_{*2}(r) \right)$. To link these eigenvalues together, compute the enveloping curve 
$$ 
\frac{d}{d\, r} \left[W_{comp}(r)-B(r) \right]=0
$$ 
of the $r$-dependent family
of bounds and obtain the system
for $k_{*1}(r) , k_{*2}(r)$:
\begin{eqnarray}
 k_{*1}(r)= B - \frac{r}{2} \frac{d\,B}{d\,r},   \quad 
k_{*2}(r)=\displaystyle  \frac{1}{2 \,r} \frac{d\,B}{d\,r}.
 \label{k-l}
\end{eqnarray}
Pairs $k_{*1}(r) , k_{*2}(r)$ form a parametric equation for the boundary of the G-closure set. If the parameter $r$ can be explicitly  excluded from system \eq{k-l}, we obtain an explicit relation between r $k_{*1}(r)$ and $ k_{*2}(r)$.

\section{Methods for bounds  and optimal structures}\label{locpoly}

\subsection{Formulation and procedure}
The technique for the bound derivation  is described in (Nesi, 1995, ; Cherkaev, 2009). Here, we repeat the arguments of the last paper. As in the translation method (se, for example. Cherkaev, 2000, Milton, 2002) we construct lower bound using quasiaffiness of  determinant (see Morrey, 1952 and Reshetnyak, 1967 ). 
\begin{equation}
  \int_\Omega \det(\nabla U) dx= \det(E_{0}) = 2r 
\label{quasiaffiness}
\end{equation}
We rewrite \eq{en-def}  adding and subtracting $2t \det E= 2 r \, t $, where $t$ is a real parameter, $t\in \mathbb{R}$ to the right-hand side terms
\begin{equation}
W_0(E_0)=\inf_{U \in H^1_{\#}(\Omega)+E_0x} \int_{\Omega}(Tr(K E E^T)+ t \mbox{det}(E)) \,dx -t\,\det(E_0)
\label{bound01}
\end{equation}

According to Translation method, we relax point-wise differential constraints on set of minimizers $E=\nabla U$, $U\in H^1_{\#}(\Omega,\mathbb{R}^{2\times 2})+E_0x$ in the integral in right-hand side of \eq{bound01}  by replacing this set with a larger set ${\cal E}$ of $E$:
\begin{eqnarray}
{\cal E} = \left\{ E:~E\in L^2(\Omega, \mathbb{R}^{2\times 2}), ~ E\in {\cal E}_{av},
 \right\}
\label{quasi}
\end{eqnarray}

We also use an extra constraint (Alessandrini and Nesi, 2001) $\det \nabla U \geq 0$ a.e. in $\Omega$, if $ \det(E_0)\geq 0$,
that further constraints the minimizers' set: $E\in {\cal E}_+$.
\begin{equation} 
{\cal E}_{+}= \left\{ E:~E\in {\cal E}, ~~s_1^2+ s_2^2 -d_1^2-d_2^2  \geq 0~~\mbox{a.e.} ~x\in \Omega \right\}
, \quad \mbox{if } r\geq 0.
\label{Nesiineq}
\end{equation}
The constraint is imposed as in (Nesi, 1995, Cherkaev, 2009).
\begin{remark}
The bound for isotropic composites obtained in (Cherkaev, 2009)  uses a stronger inequality, which, however, coincides with \eq{Nesiineq} for the considered here case $k_3=\infty$.
\end{remark}

The relaxed problem for the lower bound does not contains differential constraints. It has the form
\begin{equation}
W_0(E_0) \ge Y(r,t) -2 r\,t \label{bound1}
\end{equation}
where
\begin{eqnarray}
Y(r,t)=\inf_{E\in {\cal E}_+} \int_{\Omega}( k(s_1^2+ s_2^2 +d_1^2+d_2^2)-t(s_1^2+ s_2^2 -d_1^2-d_2^2))\,dx  \label{Y}
\end{eqnarray}
In the next section, we explicitly compute $Y(r,t)$. Inequality
\eq{bound1} gives a $t$ dependent family of lower bounds on $W(E_0)$. Let us call the best possible bound of this family $B(r)$. We have
\begin{equation}
W_0(r) \ge B(r), \quad B(r)=\max_{t} (Y(r,t)  -2 r\,t ) \label{b}
\end{equation}

Using \eq{k-l}, we find the bounds  for $K_*$, or the bounds for G-closure. The constraint  \eq{Nesiineq} could be slack or strict in different regions $\Omega_i$, depending on the  problem's parameters. If inequality \eq{Nesiineq} is slack everywhere, the procedure coincides with polyconvexification and gives the  conventional translation bound.

\subsection{Calculation of the translated energy $Y$ in the materials}
\label{str.min}
We  split the integrals into a sum of integrals over $\Omega_i$ and minimize each term independently. This is possible  because the differential constraint on the field is relaxed so that the boundary conditions between the fields in neighboring domains are neglected. Introduce the averages of the fields over $\Omega_i$,
\begin{eqnarray}
S_{ij} &=&  \frac{1}{m_i}\int_{\Omega_i} s_j(x) \, dx , \quad  
D_{ij}=\frac{1}{m_i}\int_{\Omega_i}  d_j(x)  dx , ~j=1,2.
\label{inner3a} 
\end{eqnarray}
and recall that the field in $ \Omega_3$ is zero: 
$s_j^2(x)=d_j^2(x)=0,~\forall x \in \Omega_3 ~\Rightarrow~S_{3j}=D_{3j}=0$.
Then  \eq{Y} becomes
\begin{eqnarray}
Y(r,t)= \min_{ S_{ij}, D_{ij} \in \cal{S} }
 \left(V_1+V_2 \right),
\label{b02}
\end{eqnarray}
where 
\begin{eqnarray}
 {\cal S}: ~  \left\{m_1 S_{11}+ m_2 S_{21}= S_{01},  ~m_1 S_{12}+ m_2 S_{22}= 0 \right.\nonumber , \\
 \left. m_1 D_{11}+ m_2 D_{21}= D_{01} , ~ ~m_1 D_{12}+ m_2 D_{22}= 0 \right\}
 \label{avg}
\end{eqnarray}
and
\begin{eqnarray}
 V_i  =\inf_{s^2 \ge d^2} \int_{\Omega_i} \left( (k_i+t) s^2+  (k_i-t) d^2 \right) dx, \quad i=1,2 \label{inner-pr}, \\
s^2=s_1^2+s_2^2,~~ d^2=d_1^2+d_2^2 \nonumber 
\end{eqnarray}
Here, $V_i$ are functions of $S_{ij},D_{ij}$, which we compute now

\paragraph{Structure of minimizers}
Depending on the sign of $k_i-t$, algebraic expressions for $V_i$ and corresponding minimizers $s(x),d(x)$ vary. We recognize three cases:

(a)  $k_i-t>0$ is true.
All terms in the integrand of $V_i$ are convex.  Applying Jensen's inequality to each of them, we have:
\begin{equation}
V_i=m_i(k_i+t)(S_{i1}^2 + S_{i2}^2)+m_i(k_i-t)(D_{i1}^2 + D_{i2}^2) \label{V1}
\end{equation}
The minimum is achieved when fields $s,d$ are constant:
$$
s_j(x)=S_{ij}~,d_j(x)=D_{ij}~~  \mbox{a.e.} ~x\in \Omega_i, ~~j=1,2
$$

(b)  $k_i-t <0$ is true. In this case, the $d$ terms are concave, however the inequality holds $(k_i-t)d^2 \ge (k_i-t)s^2$.  Replacement of the second term in the right-hand side in \eq{V1} by $(k_i-t)s^2$ decreases the integral value. We obtain:
\begin{equation}
V_i= \inf \int_{\Omega_i} 2k_is^2 \,dx= 2k_i  m_i(S_{i1}^2 + S_{i2}^2) \label{V2}
\end{equation}
where infimum becomes minimum again if minimizers $s,d$ are as:
\begin{equation}
s_j=S_{ij},~ d^2=s^2~ \mbox{a.e.} ~x\in \Omega_i
~ ~\mbox{or} ~~ s_j=S_{ij},~ d^2=S_{i1}^2 + S_{i2}^2=S_i^2~ \mbox{a.e.} ~x\in \Omega_i \label{min-nonconv}
\end{equation}
The minimizers $d_1(x)$ and $d_2(x)$ that correspond to the minimum assume the representations 
\begin{equation}
d_1(x) =  S_i\cos(\theta(x)), \quad d_2(x) = S_i\sin(\theta(x)), ~ \mbox{a.e.} ~x\in \Omega_i
\end{equation}
 where $\theta(x)$ is an arbitrary function. $\theta$ does not affect the value of $V$ (the bound \eq{V2} is independent of $D_i$) but the averages $D_{i1}$ and $D_{i2}$ depend on $\theta$, they vary in  a disk 
\begin{equation}
D_{i1}^2 + D_{i2}^2 \leq \left(S_{i1}^2 + S_{i2}^2\right). 
\label{av-d}
\end{equation}
All points in the  disk correspond to some $\theta(x)$, the inequality  becomes equality  when $\theta$ is constant  a.e. $ x\in \Omega_i$.

Notice that 
\begin{equation}
\det e=s_1^2+s_2^2-(d_1^2+d_2^2)=0, ~ \mbox{a.e.} ~x\in \Omega_i
\label{det0}
\end{equation}
In this case, constraint \eq{Nesiineq}  play a crucial role because of the non-convexity of term $(k_i-t)d^2$.

(c) If $k_i-t=0$ is true, second term disappears, $V_i$ becomes:
\begin{eqnarray}
V_i=\inf \int_{\Omega} 2k_is^2=2m_ik_i(S_{i1}^2 + S_{i2}^2) \label{V3}
\end{eqnarray}
which is similar to the previous case. However the minimum is achieved independently of $d$-fields that can vary  in $\Omega_i$ keeping the relations
$ d^2\le s^2 ~ a.e. x \in \Omega_i $, and $s$-components are constant, $s_j=S_{ij},~a.e. x \in \Omega_i $.

\begin{remark}
Accounting for inequality constraint,  $V_i$  can be written as
\begin{equation} 
V_i= \hat{V}_i  = \left\{\begin{array}{ll}
\inf_{s^2, d^2 \in {\cal E}} 
\int_{\Omega_i} \left( (k_i+t) s^2+  (k_i-t) d^2 \right) dx & \mbox{ if } s^2> d^2 \\
+\infty & \mbox{ otherwise} \end{array} \right. 
\end{equation}
 The integrand of $\hat{V}_i(s,d)$ is a nonconvex saddle-type function of $s$ and $d$. Due to the constraint, it is nonnegative and  grows quadratically.   One can check that expression for $V_i$ in \eq{V2} is the proper part of the convex envelope of the integrand in $\hat{V}_i$. The envelope is supported by the minimizers \eq{min-nonconv}.
\end{remark}

\paragraph{Summary} The bound is derived in following steps.
 \begin{enumerate}
\item Arbitrary fixing translation parameter $t\in [0, \infty)$, conductivities $k_1$ and $k_2$, and average field components $S_{ij}$ and $D_{ij}$, find the lower bound \eq{inner-pr} of translated energy $V_i(t, k_1,k_2, S_{ij}, D_{ij})$ in subregions $\Omega_i$  (see above).
\item Fixing $t$ and loading $E_0(r)$, minimize the energy of the cell with respect to $S_{ij}, D_{ij}$ and find optimal $S_{ij}(t, r, k_1, k_2, m_1, m_2)$ and  $D_{ij}(t, r, k_1, k_2, m_1, m_2) $,  and the bound  $Y(t, r, k_1, k_2, m_1, m_2)$ for translated energy.
\item Find optimal $t$ and energy bound $B(r, k_1, k_2, m_1, m_2)$, using \eq{bound1}.
\item Express energy bound through effective tensor $K_*$ and find equation \eq{k-l} for the G-closure boundary $ G(K_*, k_1, k_2, m_1, m_2)=0$.
\end{enumerate}

\subsection{Optimal fields and optimal laminates}\label{s:th4}

\paragraph{Method for finding matching structures}
 The exactness of obtained bounds is checked  by finding a matching laminate micro-geometries that realize them; this is done regularly as in (Albin et al, 2007a, Cherkaev, 2009). In this section, we explain the method and optimal three-material structures.
The method  assumes that obtained bounds are exact so that the range of minimizers-fields inside each material is known. The exactness of the bound is  demonstrated by constructing a sequence of multi-rank laminates, such that the field inside each phase strongly tend to fields in optimal ranges; consequently,  the energy converges to the  bounds, see for example (Conti et al,, 2003, Albin et al., 2007b)
In the derivation of the bound, the ranges of optimal fields are not restricted by compatibility conditions \eq{bound-cond},  but are determined solemnly from the optimality of a relaxed problem \eq{bound1}. The optimal laminate structure is a rank-one path between the optimal fields computed from sufficient conditions. If such path exists, the bound is realizable, and therefore optimal. 
 
A laminate with tangent $\tau $ can connect two fields $e_1$ and $e_2$ if they are {\em rank-one compatible} or $ \det(e_1-e_2)=0$.
In particular, laminates with tangent $\tau=(1, 0)$ can connect two fields 
\begin{eqnarray*}
e_1= \pmatrix{ a_1 & 0 \cr 0 & b_1} \quad \mbox{and } e_2= \pmatrix{ a_2 & 0 \cr 0 & b_2} eigenvector
\end{eqnarray*}
if $a_2=a_1$. The average field in the laminate is 
\begin{eqnarray}
e_{12}= \pmatrix{  a_1 & 0 \cr 0 & c\,b_1+ (1-c) b_2} \label{lm1}
\end{eqnarray}
where $c\in [0, 1]$ is 	the volume fraction of the first material. Here,  we consider only orthogonal laminates, with $\tau$ codirected with one of the eigenvector of $e$. The eigenvectors of laminates are the same. In the eigenvalues plane, each field is represented by a point vector $(a, b)$. The compatible fields $ e_1=(a_1, b_1)$ and   $ e_2=(a_2, b_2)$  have a common first coordinate $a_1=a_2$, 
and the set of laminates with all volume fractions is represented by an vertical interval $e_{12}=[a_1,  c\,b_1+ (1-c) b_2]$,  $ c\in [0, 1]$. 
Similarly, the layers with the tangent $(0, 1)$ connect fields with  fields $ e_1=(a_1, b)$ and   $ e_2=(a_2, b)$, and the   set of laminates with all volume fractions is represented by a horizontal interval $[c\,a_1+ (1-c) a_2, b]$, $ c\in [0, 1]$. 

\paragraph{Example 1. Second rank laminate (T-structure)} The second-rank laminate can connect three fields 
\begin{eqnarray*}
e_1= (a_1,  b_1), ~~ e_2= (a_2, b_2), ~~ e_3=(a_3, b_3), \quad \mbox{if }a_1=a_2, ~~b_3\in [b_1,b_2]
\end{eqnarray*}
First, we connect fields $e_1$ and $e_{2}$ by laminates as in \eq{lm1} and choose $c=c_0$ so that 
$$
c_0\,b_1+ (1-c_0) b_2= b_3, $$
then laminate $e_{12}$, \eq{lm1} becomes compatible with $e_3$ and can be joined by an orthogonal laminate with tangent $\tau_h=(0, 1)$. 
We call the resulting structure $ L(12,3)$ pointing out that fields $e_1$ and $e_2$ are joined by laminates, then the structure is rotated $90^\circ$ (the rotation is shown by comma) and laminated with $e_3$. This rank-one connection of $e_3$ with the laminate from $e_1$ and $e_2$  is possible if the relative volume fractions in that laminate are specially fixed. 
The volume fraction 
of the third material can be arbitrary. 
The average field $e_{12,3}$ in the structure is simply a sum $e_{12,3}=m_1 e_1+ m_2 e_2 + m_3 e_3$. 
\begin{remark}
Strictly speaking, the second-rank laminate is a sequence of structures, the corresponding energies converge to the calculated values when the ratio of laminates scales goes to zero,
see for example (Conti, 2003).
\end{remark}

\paragraph{Rank-one path through given field ranges.}
The described procedure can be iteratively applied to find an optimal laminate that links fields $e_i$ in the given ranges. The  process is continued until all materials are joined together and the average field of the whole structure is equal to the required value $e_0$. In the eigenvalues plane, the process of formation of an optimal structure  is represented by a graph that joints the permitted  values $e_i$ by  either horizontal or vertical intervals, and the last interval must pass through the given point $e_0$. 
The volume fractions that determine the positions of the fields within intervals are not shown in this graph. They must be computed separately, and the constraints on volume fractions determine the domain of applicability of the graph. Usually, the graph (and corresponding optimal microstructures) is not unique.

The next two examples demonstrate the earlier obtained optimal three-phase laminates which are parts of the whole picture presented here. The structures were originally obtained without the assumption $k_3=\infty$, but we keep it here.

\paragraph{Example 2: Optimal structures for the Translation bounds}

Following (Albin et al. (2007)), we construct anisotropic optimal structures that correspond the following sufficient conditions for the fields in materials 
\begin{equation}
e_1= [ \lambda a_1, (1-\lambda) a_1 ],  \quad  e_2= [a_2, a_2], \quad e_3=[0, 0]
\label{tr-fields}
\end{equation}
where $\lambda$ is an arbitrary real parameter, and $a_1, ~a_2$ are constants. Field $e_1$ in the first material is underdetermined, and the field in second material is proportional to a identity matrix. These relations are derived from Translation sufficient condition (Albin at al, 2007a).  We show here the rank-one path (laminate microstructure) that realizes the condition.

\begin{figure}
\begin{center}
\includegraphics[scale=.5]{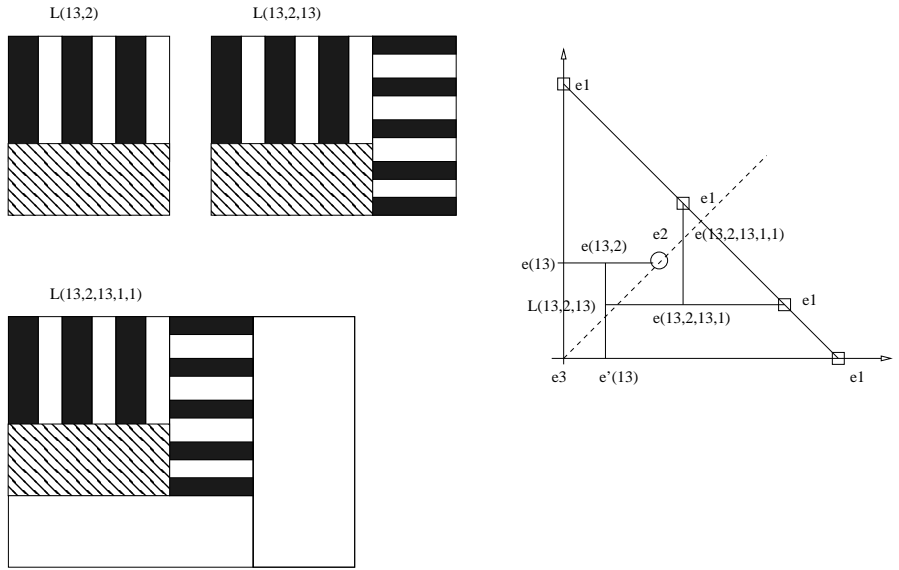} 
\caption{Scheme of fields in a structure that realizes the translation bound} 
\label{f:lm2}
\end{center}
\end{figure}
The permitted by translation bounds fields in materials one and three are in rank-one contact if we take $\lambda =0$ or $\lambda =1$ in \eq{tr-fields}. Joining by laminates certain amounts of materials  one (where we take $\lambda=0$) and three, we obtain  laminate (interval $A_{10} $) and call it $L(13)$. Fixing the relative volume fractions of materials in the laminate, we arrive    at the point $[0, e_{{13}}]= [0, a_2]$  that is compatible with material two, field $[a_2, a_2]$. Then  a second-rank laminate $L(13,2)$ is formed by adding this material in an orthogonal layer, its (average) field is represented by a point $e_{12,3}$ of the interval $[e_{{13}}, e_2]$. 
Then, we join materials one and three (where we take $\lambda=1$, see \eq{tr-fields}) in an orthogonal laminate,  choose the  relative volume fractions so that the average field in the laminate is compatible with the filed $e_{12,3}$, and laminate it with the structure $L(13,2)$ obtaining structure $L(13,2,13)$. 
Assume that all amounts of materials two and three are used to build the structure, but there remains an extra amount of material one. To add this material, we choose an appropriate value of $\lambda$ in \eq{tr-fields}, laminating the structure $L(13,2,13)$ with the first material and forming structure $L(13,2,13,1)$, then  laminating it again in an orthogonal direction, obtaining structure $L(13, 2, 13, 1, 1)$. 
The parameters of the last two laminations must be chosen to reach a prescribed average field. 

The limits on relative volume fractions or on the attainable by this construction average field can be transformed  into range of applicability, see (Albin et al. (2007)). It turns out  that  volume fraction $m_1$ must be above a threshold $g_m(r)$, $ m_1\geq g_m(r) $  and the rate of anisotropy $r$ must be large enough, $r\geq g_r$, too.

\paragraph{Example 3: Isotropic optimal three-material structures}

Similar technique was used in (Cherkaev, 2009) to determine optimal structures of  isotropic composites $(r=1)$ beyond the applicability of  translation bound, $m_1<g_m(1)$.  
The optimal isotropic structures as similar to the structures shown below in Figure \ref{f:str}, regions A, B, D1). It is shown   that one of three types of laminates are optimal, depending on the range of volume fraction $m_1$. There are two critical values $0< m_{00} < m_0 <1$ that determine the topology of optimal structures:
If $ m_1\in [m_{0} ,\, m_1]$, the described above structures $L(13,2,13,1,1) $  are optimal, if $ m_1\in [m_{00} ,\, m_0]$, the optimal structure degenerate into $L(13,2,13) $, and if $ m_1\in [0, \,m_{00}]$, the optimal structure becomes $L(123,2,123) $.
The regions of applicability of both structures match the regions of applicability of the bounds. 
\begin{remark}
Below in Section \ref{s:structA}, we suggest different than  $L(123,2,123) $ optimal structures for the case $m_1\leq m_{00}$, that are more convenient for description of anisotropic composites. This variation is possible because the rank-one path between fields in optimal ranges is nonunique. 
\end{remark}

\section{Results}\label{s:results}

In this section, we summarize the results. Their derivation is shown in the two following sections.
\paragraph{Bounds for the energy and minimizers.}

\begin{figure}[htbp]
\begin{center}
\includegraphics[scale=0.5]{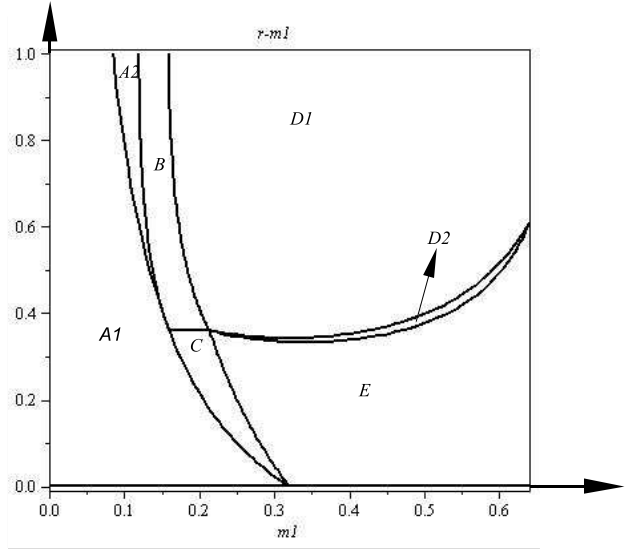} 
\caption{{\bf Regions A-E of multifaceted boundary for the energy in an anisotropic field in the $r, m_1$ plane.}}
\label{f0}\label{f:allregions}
\end{center}
\end{figure}

The analysis in Section \ref{s:bounds} demonstrates that the optimal bound for energy is a multifaceted surface: it is expressed by different analytic formulas in different domains. These domains are conveniently presented in the parameters plane $r, m_1$, Figure \ref{f0} where the parameters were set as $k_1=1, k_2=3, m_2=.5$ The dependence on $m_2$ is not shown on the figures. Variation of $m_2$ leads to variation of the shape of the regions above but does not change their topology.  It turns out that there are five nontrivial cases (Regions A, B, C, D, and E), Figure \ref{f0}, depending the values of $t_{opt}(k_1,k_2, m_1, m_2)$ and $r$. Region D corresponds to the known translation bound (Milton and Kohn, 1988), optimal structures in region D1 have been found in (Albin et al. (2007)), the isotropic structures were found in (Gibiansky and Sigmund, 2000). The upper interval $r=1$ correspond to isotropy of $K_*$, the isotropic bound have been derived in (Nesi 1995), and optimal structures have been found in (Cherkaev, 2009). Other bounds and structures  are new. 

Boundaries between regions in Figure \ref{f0} are calculated, the expressions are shown in Table \ref{t22}. The division between A1 and A2 and between D1 and D2 are based on structural attainability, the bounds are given by the same expressions. 

\begin{figure}[htbp]
\begin{center}
\includegraphics[scale=0.5]{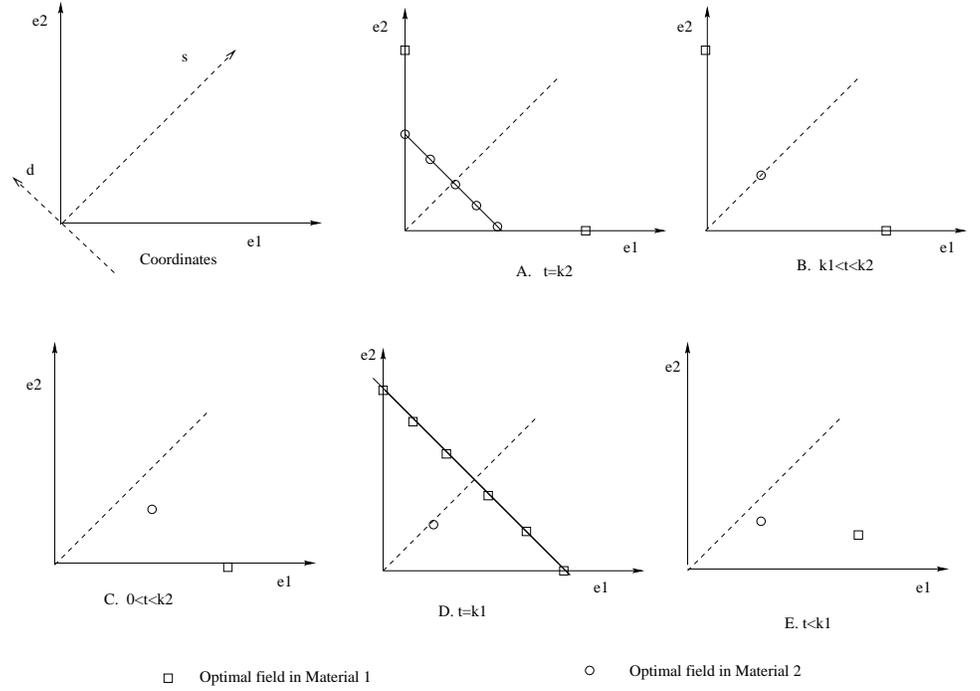} 
\caption{{\bf The eigenvalues of the gradient field- minimizers, according to the bounds. The equilibrium condition is not assumed.
}}
\label{f1}
\end{center}
\end{figure}

  \begin{table}[htdp]
\caption{Boundaries between regions}
\begin{center}
\begin{tabular}{ |l|l|l|l|l|l|l|l| l|}
\hline 
Boundary &A1A2& A2B &BD& A1C,& CE & BC& D1D2& D2E\cr
\hline
Formula &\eq{boundA1A2}  & \eq{boundAB} &\eq{boundBD}& \eq{bound-AC}& \eq{bound-CE}& \eq{boundBC} &\eq{psi-d} & \eq{psi-E} \cr
\hline
\end{tabular}
\end{center}
\label{t22}
\end{table}
 Table \ref{t21} summarized the expressions for optimal energy bound in different regions, expressions for G-closure boundary, and optimal laminates that realize the boundary or best approximate it. 
 \begin{table}[htdp]
\caption{Formulas for energy bounds and G-closure, and type of optimal structure or the best approximate}
\begin{center}
\begin{tabular}{ |l|l|l|l|l|  }
\hline 
Region &   Energy  & G-closure  & Structure & Exact? \cr
\hline
A1 &  \eq{wa} &  \eq{t=k2bound} & L(13,2,13,2) & yes\cr
A2 &  \eq{wa} &  \eq{t=k2bound} & L(123,2) & yes\cr
B &  \eq{enB} & \eq{gcl-par}  & L(13,2,13) & yes\cr
C &  \eq{en-C} & \eq{gcl-C}  & L(13,2) & yes\cr
D1 &  \eq{bound-d} & \eq{transl-bound}  & L(13,2,13,1,1) & yes\cr
D2 &  \eq{bound-d} & \eq{transl-bound}  & L(13,2,1) & not known\cr
E &  \eq{we} &   & L(13,2,1) & no\cr
\hline
\end{tabular}
\end{center}
\label{t21}
\end{table}%
Table \ref{t1} and Figure \ref{f1} show the details of optimal parameters and minimizers in different regions.
 
 \begin{table}[htdp]
\caption{Character of minimizers}
\begin{center}
\begin{tabular}{|l |l|l| l | l | l|}
\hline 
Region &  $t_{opt}$ & $d(x) $ in $\Omega_1$ &  $d(x) $ in  $\Omega_2$  \cr
\hline
A  & $t=k_2$& $ d_{11}^2+d_{12}^2 = S_{11}^2 $  & $d_{21}^2+d_{22}^2 \le S_{21}^2		$  \cr
B &   $ t\in(k_1,k_2) $ &  $d_{11}^2+d_{12}^2 = S_{11}^2   $  &  $d_{2j}=0 $  \cr
C &    $t\in (k_1,k_2)$&  $d_{11}=S_{11},~d_{12}=0 $  & $d_{21} =\mbox{const} >0, ~d_{22}=0$  \cr
D1 &    $t =k_1$ & $d_{11}^2+d_{12}^2 \le S_{11}^2  $ & $d_{21} =0~d_{22} =0 $ \cr
D2 &    $t =k_1$ & $d_{11}^2+d_{12}^2 \le S_{11}^2  $ & $d_{21} =0~d_{22} =0 $   \cr
E &     $t\in (0, k_1)$ & $d_{11}= \mbox{const}>0, ~d_{12}=0$  &  $d_{21}= \mbox{const}>0, ~d_{22}=0$   \cr
\hline
\end{tabular}
\end{center}
\label{t1}
\end{table}%

\begin{figure}[htbp]
\begin{center}
\includegraphics[scale=0.5]{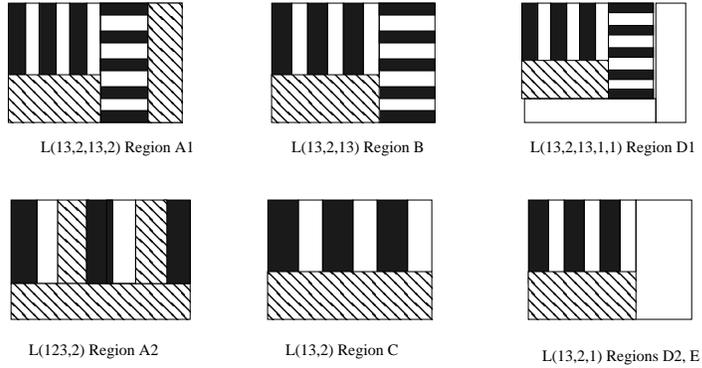} 
\caption{{\bf 
Cartoon of optimal structures in regions A -D1 and the presumed optimal structure in region E.
}}
\label{f:str}
\end{center}
\end{figure}

\paragraph{Three special points} In the map of the regions, notice two points where several region meet.
The  four regions A1, A2, B, and C meet in the point P1:
\begin{eqnarray}
r=m_2, \quad m_1=\frac{k_1(1- m_2)}{2k_2}
\end{eqnarray}
The four regions B, C, D, and E meet at the point P2
\begin{equation} r=m_2,  \quad m_1=\frac{k_1(1-m_2)}{k_1+k_2}.
\end{equation}
At the line (P1, P2) the field $E_2$ is constant and proportional to the identity matrix, and below this line the proportionality is lost. 

The three regions A2, C and E meet at the point P3:
\begin{equation}
 r=0,  \quad m_1=\frac{k_1(1-m_2)}{k_2}. 
\end{equation}
At this point, the L(12,3) structure has the same conductivity in $x_1$ direction, as a simple laminate, see (Cherkaev and Gibiansky, 1996).

\paragraph{G-closure boundaries.}  

\begin{figure}[htbp]
\begin{center}
\includegraphics[scale=0.6]{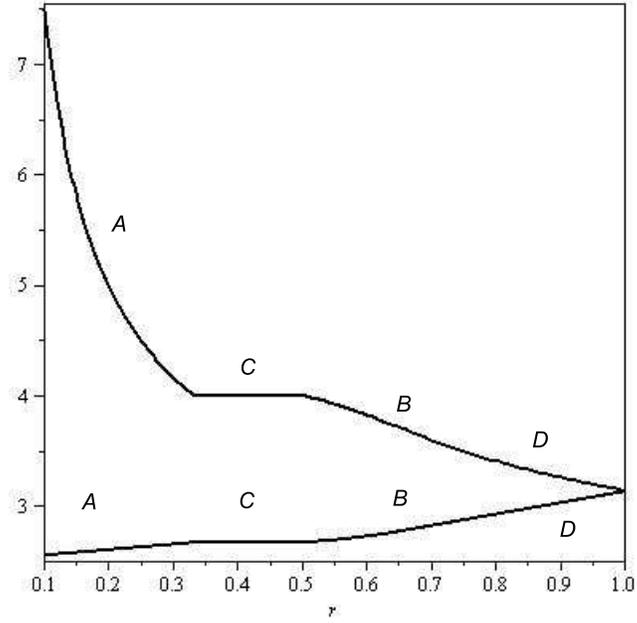} 
\caption{{\bf The eigenvalues of effective tensor $K_*$ at the G-closure boundary in dependence of $r$.}}
\label{f:g-cl1}
\end{center}
\end{figure}

The results for the G-closure boundary follow from the energy bounds, the expression for G-closure boundary are summarized in Table \ref{t21}.  In Figure \ref{f:g-cl1}, the eigenvalues $k_{*1}(r), ~k_{*2}(r)$ are shown that form the G-closure boundary for the values $k_1=1, ~k_2=2,~m_1=0.15, ~m_2=0.5$. Notice, that decreasing values of $r$ correspond to regimes D, B, C, and A (see Figure \ref{f0}). The horizontal line in the graph represent region C, because both eigenvalues are constant independent of $r$.  

\begin{figure}[htbp]
\begin{center}
\includegraphics[scale=0.5]{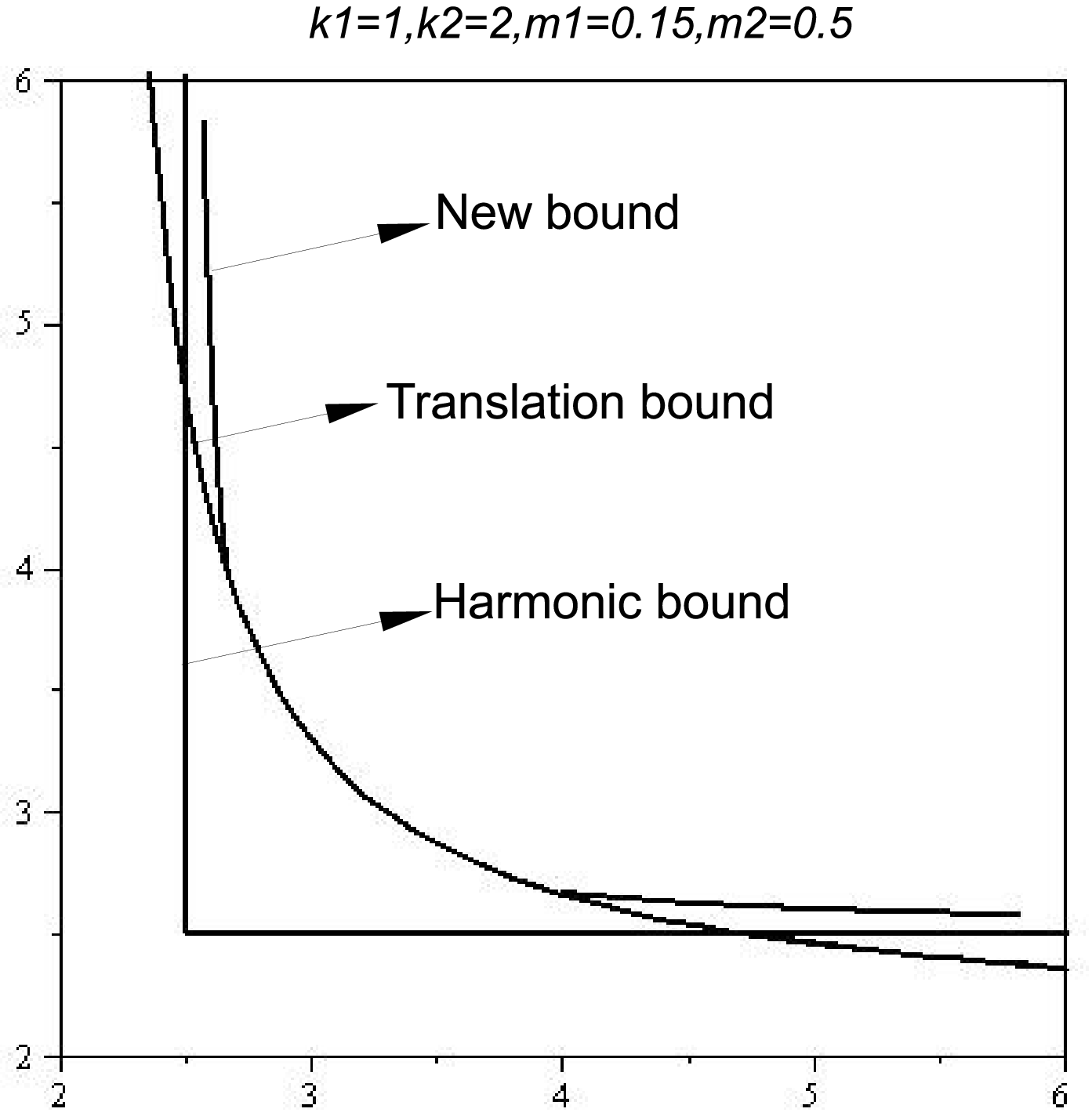} \includegraphics[scale=0.6]{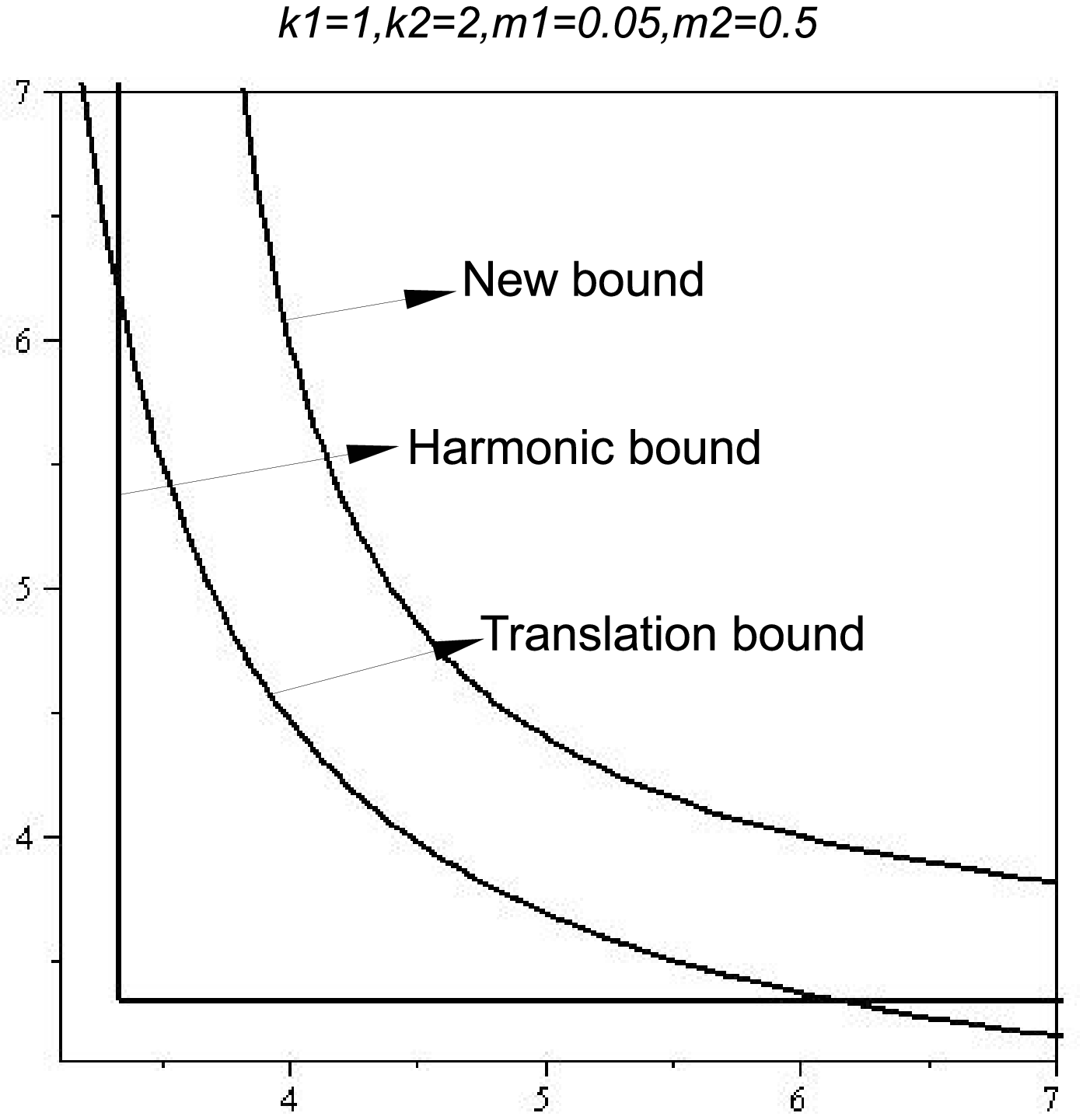} 
\caption{{\bf The G-closure boundary and its estimates (harmonic and translation bounds). $k_1=1, k_2=2, ~m_2=0.5$
Left: $m_1=0.15$ (Regions D, B, C (sharp bend point), A). ~ Right: $m_1=0.05$ (Region A)  }}
\label{f:g-cl2}
\end{center}
\end{figure}

Figure \ref{f:g-cl2} shows the G-closure boundary, parameter $r$ is excluded. The graphs show the result of comparing the bounds obtained here with the translation and harmonic mean bounds for two values of $m_1$.   

\section{Derivation of the bounds}\label{s:bounds}

Solving minimization problem \eq{b02} for fixed $t$, we notice that the algebraic form of  $Y(r,t)$ depends on sign$(k_1-t)$ and sign$(k_2-2)$. 
There are five cases to analyze: $ t<k_1, ~t=k_1$ (translation bound), $k_1<t<k_2, ~t=k_2$, and $t>k_2$.  After the solution $Y(r,t)$ of \eq{b02} is obtained, the bounds are derived by  maximizing the bound with respect to $t$, and G-closure bound follows. 

\subsection{Intermediate value $\mathbf{t\in (k_1,k_2)}$} 

\paragraph{Study of minimization problem  \eq{b02}}
If  $k_1< t< k_2$, $V=V_1+V_2$ is computed using   \eq{V1}, \eq{V2}
\begin{equation}
V=2m_1k_1(S_{11}^2 +S_{12}^2)+m_2(k_2+t)(S_{21}^2+S_{22}^2 ) + m_2(k_2-t)( D_{21}^2+D_{22}^2) \label{ff5}
\end{equation}
We compute $Y(t, r)$ from the relations (see \eq{avg}, \eq{av-d})
\begin{eqnarray}
&~&Y(t, r) =\min_{ S_{11},~S_{12},~S_{21},~S_{22}, ~D_{21},~D_{22} }V  \nonumber \\
& ~& \mbox{subject to: } \nonumber \\
& ~& m_1 S_{11}+ m_2 S_{21}= S_{01}=\frac{1+r}{\sqrt{2}}, \quad m_1 S_{12}+ m_2 S_{22}= 0, \label{cs-S} \\
& ~& m_1 D_{11}+ m_2 D_{21}= D_{01}=\frac{1-r}{\sqrt{2}}, \quad m_1 D_{12}+ m_2 D_{22}= 0,  \label{intt-const}\\
& ~& D_{11}^2+ D_{12}^2 \leq   S_{11}^2+ S_{21}^2  \nonumber \\
\end{eqnarray}
The problem for $Y$ is a standard finite-dimensional constrained optimization problem for the averages $D_{ij}$ and $S_{ij}$. The analysis of optimality Karush - Kuhn -Tucker conditions leads to the following optimal values:
$$ S_{21}=S_{22}=D_{22}=0 $$ 
Depending on whether or not the last constraint \eq{av-d} is satisfied as an equality or as a strict inequality for optimal values of  $D_{ij}$ and $S_{ij}$. We recognize two cases: 
\begin{enumerate}
\item 
(Case B): If $\ {m_1} \, S_{11} \geq D_{01} $ then in the optimal values are: $\displaystyle D_{11}=\frac{D_{01}}{m_1}$ and $D_{21}=0$.

\item (Case C):
If $0 \leq m_1 D_{11} < D_{01}$  even when $D_{11}$ reaches its maximum $D_{11}=S_{11}$, that is equivalent to $ \displaystyle D_{11}=S_{11} < \frac{D_{01}}{m_2}$,  
then  
\begin{eqnarray}
D_{21}=\frac{D_{01}-m_1S_{11}}{m_2}>0 ~\mbox{ and } ~ D_{11}=S_{11}
 \label{D21_B}
\end{eqnarray}
In this case, the $d$-minimizer is constant $d_1(x)=D_{11}=S_{11}, ~d_2(x)=0,~~\mbox{a.e.}~ x \in \Omega_1$.
\end{enumerate}
Now we compute minimum $Y(r,t)$ of $V$ and  optimal fields   $D_{ij}$ and $S_{ij}$ for these two cases as functions of $t$ and given parameters $r$, $m_i$, and $k_i$.

\paragraph{Case B} If $D_{21}=0$ holds,  
\eq{ff5} becomes:
$$
V=2 m_1 k_1 S_{11}^2  + m_2 (k_2+t)S_{21}^2 
$$
The minimum of $V$ subject to the constraint \eq{cs-S} can be found using the standard Lagrange multiplier procedure:
\begin{equation}
Y_B(t, r) = H_0S_{01}^2= \frac{(1+r)^2}{2} H_0
\label{70}
\end{equation} 
and the optimal $S_{i1}$ are:
\begin{eqnarray}
S_{11}= \frac{1+r}{2\sqrt{2} k_1}H_0, \quad S_{21}= \frac{1+r}{\sqrt{2}(k_2+t)}H_0, 
\label{avrB}
\end{eqnarray}
where
$$
H_0= \left(\frac{m_1}{2k_1}+\frac{m_2}{k_2+t}\right) ^{-1}
$$
and point-wise minimizers satisfy:
\begin{eqnarray*}
d_1^2+d_2^2=S_{11}^2,~ s_1=S_{11}, 
~s_2=S_{12}=0,~~\mbox{a.e. in}~ \Omega_1 \\
d_1=D_{21}=0, d_2=D_{22}=0, s_2=S_{22}=0, s_1=S_{21}, ~\mbox{a.e. in}~ \Omega_2 
\end{eqnarray*}
which implies that:
\begin{eqnarray}
\mbox{Tr}(e)&=&\mbox{constant}, \quad  \det(e)=0 ~~\mbox{a.e. ~in} ~\Omega_1, \nonumber \\ 
& & e=\alpha I, ~\mbox{a.e. in} ~\Omega_2 \label{opt_B}
\end{eqnarray}
Bound $B_B$ can be found by solving:
$$ B_B= \max_{ t \in (k_1,k_2)}( Y_B(t,r)- 2tr)$$
There are two critical values of $t$:
\begin{eqnarray}
{t_{cr}}_1 &=& -\frac{r\,m_1k_2+2r\,m_2k_1-k_1\sqrt{rm_2}(1+r)}{r\,m_1}  \label{tcr1}\\
{t_{cr}}_2 &=& -\frac{r\,m_1k_2+2r\,m_2k_1+k_1\sqrt{rm_2}(1+r)}{r\,m_1} \nonumber
\end{eqnarray}
and it can be checked that maximum is achieved at ${t_{cr}}_1$, therefore $t_{opt}={t_{cr}}_1$. Substitute it back into $B_B$ and we get energy bound:
\begin{equation}
B_B=  \frac{k_1(1+r-2\sqrt{rm_2})^2}{2m_1}+r\,k_2
\label{enB}
\end{equation}

\paragraph{$G$-closure boundary} After the expression for $B_B$ is obtained, a parametric representation for  $k_{*1}$ and $ k_{*2}(r)$ at the boundary of $G$-closure are found using (\ref{k-l}): 
\begin{eqnarray}
 k_{*1}=K(r) \quad k_{*2}=K\left(\frac{1}{r}\right)
\label{gcl-par}
\end{eqnarray}
where
\begin{eqnarray*}
K(r)=\displaystyle \frac{k_1 \left(1-\sqrt{rm_2} \right) \left( 1 +r - 2\sqrt{rm_2}\right)}{m_1} + rk_2
\end{eqnarray*}
Eigenvalues $k_{*1}(r)$ and $k_{*2}(r)$ are linked at the boundary of G-closure through parametric equation
\begin{eqnarray*}
\frac{1}{k_{*1}-{t_{cr}}_1(r)}+ \frac{1}{k_{*2}-{t_{cr}}_1(r)} =\frac{2}{H_0- 2 {t_{cr}}_1(r)}
\end{eqnarray*}
We show below in Section \ref{s:struct} that bound $B_B$ is achievable by laminate $L(13,2,13)$.

\paragraph{Case C} If $D_{21} >0$, $V$ is expressed as:
\begin{equation}
V=2 m_1 k_1 S_{11}^2  + m_2 (k_2+t)S_{21}^2+ m_2(k_2-t)D_{21}^2
\end{equation}
substitute $D_{21}$ given by \eq{D21_B} into $V$ and also express $S_{21}$ in terms of $S_{11}$ using \eq{cs-S} we get a quadratic function of a single variable $S_{11}$:
$$
V=2 m_1 k_1 S_{11}^2+ (k_2+t)\frac{(S_{01}-m_1S_{11})^2}{m_2} +(k_2-t)\frac{(D_{01}-m_1S_{11})^2}{m_2}
$$
Differentiate it with respect to $S_{11}$ and we can find the optimal $S_{11}$ and $Y(t,r)$
\begin{eqnarray}
S_{11} &=& \frac{S_{01}\left(k_2+t\right)}{2\tilde{k}}+\frac{D_{01}\left(k_2-t\right)}{2\tilde{k} } 
=\frac{(k_2+rt)}{\sqrt{2}\tilde{k}} 
\label{interm3} 
\\
 Y_C &=& 2\frac{- m_1r^2 t^2+ 2 m_2\left(k_1-\tilde{k} \right) r \,t 
+   m_2 k_1 k_2  +  \tilde{k} k_2 r^2}{2m_2\tilde{k}}  
\nonumber
\end{eqnarray}
 where 
\begin{equation}
 \tilde{k} = m_1 k_2+ m_2 k_1
\label{tilde}
\end{equation}
Differentiating $Y_C(r,t)-2rt$ with respect of $t$, we find the optimal value $t_{opt}$:
\begin{equation}
t_{opt}=\frac{\left(k_1-\tilde{k} \right)m_2}{rm_1} \label{topt2}
\end{equation}
and the bound
\begin{equation}
B_C=  \frac{ \left( 1-m_2\right)^2 k_1+ m_1 m_2 k_2}{2m_1} +\frac{k_2}{2m_2} r^2 
\label{en-C}
\end{equation}
\paragraph{G-closure boundary} 
According to (\ref{k-l}), we have:
\begin{eqnarray}
k_{*1}= \frac{ \left( 1-m_2\right)^2 k_1+ m_1 m_2 k_2}{m_1}, \quad
k_{*2}= \frac{k_2}{m_2} \label{gcl-C}
\end{eqnarray}
The whole region corresponds to a single point at G-closure boundary. 
\begin{remark}
In optimal two-material mixtures,  the optimal two-material structure is a simple laminate when $r$ is less than a threshold value. Here, in optimal three-material mixtures, the $L(13,2)$ structure is analogous to a laminate, and it also possesses a constant fields in each subdomain $\Omega_i$. 
\end{remark}
\paragraph{Boundaries of region B and C}
The applicability of bounds is described through systems of inequalities of $r$ in terms of conductivities $k_i$ and volume fractions $m_i$. Part of inequalities is obtained by solving the constraint $k_1 <t_{opt} <k_2$.
The rest is derived from the requirement 
\begin{eqnarray*}
\frac{D_{01}}{m_1} \le S_{11}~~ \mbox{in case B} \quad
\mbox{or} ~~ \frac{D_{01}} {m_1}\ge S_{11}~~ \mbox{in case C}.
\end{eqnarray*}
The results are presented below.

{\bf{Region B}}
\begin{eqnarray}
 & & \psi_{AB} < r < \psi_{BD}, \quad r\ge m_2  \nonumber\\
 & & \psi_{AB} =m_2 \left( \frac{k_1}{\tilde{k}+\sqrt{ \tilde{k}^2-m_2k_1^2}} \right)^2 \label{boundAB}
\\
& & \psi_{BD} = m_2 \left(\frac{ 2k_1}{a+\sqrt{a^2-4m_2k_1^2}}\right)^2, \quad  a=\tilde{k}+2m_2k_1 \label{boundBD}
\end{eqnarray}

{\bf Region C}
\begin{eqnarray}
& & \psi_{AC} < r < \psi_{CD}, \quad r\le m_2  \nonumber\\
& & \psi_{AC}= \frac{m_2}{m_1}  \frac{(k_1-\tilde{k})}{k_2}, \label{bound-AC} \quad  \psi_{CE}=   \frac{ m_2}{m_1} \frac{(k_1-\tilde{k}) }{k_1} \label{bound-CE}. 
\end{eqnarray}
The analysis of constraints \eq{b02} tells that bound ${B}_C$ is valid, if  
\begin{equation}
D_{11}=S_{11}, ~~m_1D_{11}-D_{01}<0. \label{cond1}
\end{equation}
otherwise, bound $B_B$ is effective.
To work out this inequality, we substitute expression (\ref{interm3}) for optimal $D_{11} (t, D_{01}, S_{01})$  into (\ref{cond1}), account for relationship $\displaystyle D_{01}=\frac{1-r}{1+r} S_{01}$, and obtain condition 
$$ r\left(m_1 t_{opt}(r)+\tilde{k} \right)-k_1m_2
<0$$
of realizability of region C.
Substituting $t_{opt}$ (see (\ref{topt2})) into the above inequality and  simplifying, we bring this condition to the form:
{ $r-m_2 < 0 $ in Region C}.
The boundary between regions B and C corresponds to the equality
\begin{equation}
r-m_2=0
\label{boundBC}
\end{equation}

\subsection{Case $t=k_2$.}\label{s:t=k2}
\paragraph{Region A:} 
We follow the same procedure as in the previous section. First we find $V=V_1+V_2$ by evaluating $V_1, V_2$ in this case (see \eq{V2}, \eq{V3}):
$$
V=2k_1 m_1( S_{11}^2+S_{12}^2 )+ 2k_2m_1 (S_{21}^2 +S_{22}^2)
$$
$V$ is independent of $D_{1j}, D_{2j}$. We minimize $V$subject to the  only constraints $$m_1S_{1j} +m_2S_{2j} =S_{0j}, ~j=1,2$$ The constraints $$m_1D_{1j} +m_1D_{1j}=D_{0j}$$ are satisfied by point-wise minimizers  $d_j, j=1,2$ in $\Omega_i, i=1,2$, such that:
\begin{eqnarray}
 d_1^2+d_2^2 &\le & s_1^2+s_2^2,~~\mbox{and} ~~ s_j=S_{2j},\hspace{1in} a.e~ \mbox{in} ~~\Omega_2, \nonumber \\
d_1^2+d_2^2&=&s_1^2+s_2^2,~~ \mbox{and} ~~s_j=S_{1j}, \hspace{1in} a.e.~\mbox{in} ~~ \Omega_1, \label{sd_A}
\end{eqnarray} 
These minimizers exist, because $D_{0j}\le S_{0j}$ for $r\in[0,1]$. 

The minimum of $V$ is found using the standard Lagrange multiplier procedure:
\begin{equation}
S_{1j}= \frac{H_2}{2k_1} S_{0j}, \quad S_{2j}= \frac{H_2}{2 k_2} S_{0j}, \quad j=1,2
\label{avrA}
\end{equation}
or
$$
S_{11}=\frac{H_2(1+r)}{2\sqrt{2}k_1} , ~~S_{21}=\frac{H_2(1+r)}{2\sqrt{2}k_2},~~S_{12}=S_{22}=0
$$
where
$$
H_{2} = \left(\frac{m_1}{2k_1}+\frac{m_2}{2k_2}\right)^{-1}.
$$

We conclude that $S_{11}$ and $S_{21}$ satisfy the relation 
\begin{equation}
\frac{S_{11}}{S_{21}} =\frac{k_2}{k_1}   \label{Si1_A}
\end{equation}
Besides, \eq{sd_A} and \eq{avrA} implies that point-wise minimizer in $\Omega_2$ satisfies:
\begin{eqnarray}
Tr(e)&=&\mbox{constant}, ~~a.e.~\mbox{in} ~\Omega_i,~i=1,2, \nonumber \\
\det(e)&=& 0,\hspace{.5in}~~a.e.~\mbox{in}  ~\Omega_1, \label{opt_A}
\end{eqnarray}

Accounting for the value of $S_{01}$, $\displaystyle S_{01}=\frac{1+r}{\sqrt{2}},S_{02}=0$, we compute
\begin{eqnarray}
Y_A(r)&=& H_2 S_{01}^2=\frac{H_2(1+r)^2}{2}
\label{ff2}
\end{eqnarray}
Bound $B_A$ is:
\begin{equation}
B_A=\frac{H_2(1+r)^2}{2}-2k_2r \label{wa}
\end{equation}
\begin{remark}
We notice that when $t \in (k_2,\infty)$, expression for $V$ is the same as in the case where $t=k_2$ (although the pointwise $d_j,  ~ in~ \Omega_2$ is different, see Section \ref{str.min}  for details). Therefore, expression for $Y$ is the same as $Y_A$ (see \eq{ff2})and it is independent of $t$. Maximizing $Y-2tr$ with respect to $t$ to find the bound, we find that $t_{opt}=k_2$, because $Y_A-2tr$  linearly decreases with  $t$. In conclusion, if it is assumed that  $t\in (k_2,\infty)$, the optimal value is $t=k_2$ and the bound is as in  $B_A$.
\end{remark}
{\bf $G$-closure boundary} Find  $k_{*1}(r)$ and $ k_{*2}(r)$ again using \eq{gcl-par}
where
\begin{eqnarray*}
k_{*1}=K_A(r) =\frac{1}{2} (r+1) H_2  - r k_2, ~~k_{*2}=K_A(1/r).
\end{eqnarray*}
Excluding $r$ from these two equations, we find the equation describing the boundary. One can check that 
 $k_{*1}(r)$ and $ k_{*2}(r)$  are bounded as 
\begin{equation}
\frac{2}{H_2- 2 k_2}=\frac{1}{k_{*1}-k_2}+ \frac{1}{k_{*2}-k_2}
\label{t=k2bound}
\end{equation}
This relation is similar to Translation bound (region D). If  the composite is isotopic, the bounds coincide with the one found in (Nesi, 1995)
\paragraph{Region of applicability}The region $\Phi_A$ where bound $W_A$ is effective is bounded   by  inequalities $ m_1\geq 0$, $ 0\leq r \leq 1$    and by curves $\phi_{AB}$ and $\phi_{AC}$:
\begin{eqnarray*}
  r &\leq& \phi_{AB}, ~m_2 \leq r \leq 1, \label{pa1}\\
 r &\leq &\phi_{AC}  \quad 0< r \leq m_2 \label{pa2}
\end{eqnarray*}
In coordinates $m_1, r $, the region is shaped as a curvilinear pentagon, It corresponds to small values of $m_1$ and all range of $r$ and is side neighboring the region B and C.

\subsection{Case $t=k_1$ ( Translation bound)}\label{s:transl-bound}

\paragraph{Region D} This region is analogous to region B. It corresponds to the conditions $t_{opt}=k_1$ and moderate anisotropy level that corresponds to optimal values $D_{2j}=0$. The bound
corresponds to the classical Hashin-Shtrikman (1983) (for $r=1$),  and translation (Milton and Kohn, 1988), (for $r\leq 1$) bounds. The bound can be viewed as  a special case of Case B, that corresponds to special value of $t$. The expression for $V$ and averages are as in \eq{avrB} where we put $t=k_1$:

\begin{eqnarray}
S_{11} &=& \frac{S_{01}}{2k_1}H_1,  \quad 
S_{21} = \frac{S_{01}}{k_1+k_2}H_1, \nonumber \\ 
D_{11}&= &\frac{D_{01}}{m_1} \leq S_{11},~ D_{12}=0, ~ D_{2j}=0
\label{ff8}
\end{eqnarray}
where
\begin{eqnarray*}
H_1 ^ {-1}= \frac{m_1}{2k_1}+\frac{m_2}{k_1+k_2}.
\end{eqnarray*}
We compute $ Y_D=H_1S_{01}^2 $ and the bound $B_D$ is:
\begin{equation}
B_D = H_1\frac{(1+r)^2}{2}-2r\,k_1
\label{bound-d}
\end{equation}

Minimizers $s(x)$ are constant in all domains, $s_{ij}(x)=S_{ij}$. Minimizer $d(x)$ is zero in $\Omega_2$ and freely varies in $\Omega_1$  as long as the inequality $\sum_{j} d_j^2(x) \le \sum_{j} S_{1j}^2$ holds, keeping its mean  value $D_{1j}$ as in \eq{ff8}.

\paragraph{G-closure Boundary} In Region D, the calculations are similar to case $A$ with $k_2$ replaced by $k_1$ and $H_2$ by $H_1$, see (\ref{t=k2bound}). The effective conductivities are given by \eq{k-l}. Parameter $r$ can be excluded, and  the boundary of $G$-closure corresponding to the Translation bound is (compare bounds for two-component mixture Lurie and Cherkaev, 1982, Tartar, 85):
\begin{equation}
\frac{1}{k_{*1}-k_1}+ \frac{1}{k_{*2}-k_1}=\frac{2}{H_1- 2 k_1}
\label{transl-bound}
\end{equation} 

\subsection{Small values $t\in (k_1, 0]$}\label{s:small-t}

{\bf Region E} corresponds to $t_{opt} < k_1$. 
In this case,  $V_i$'s are convex functionals of $d(x)$ and the minimizers $d(x)$ and $s(x)$ are constant 
$$
s(x)_j=S_{ij}, ~d(x)_j=D_{ij} ~\mbox{ in }\Omega_i, \quad \mbox{if } 0\leq t< k_1
$$
The minimizers are computed by the same procedure as before:
\begin{eqnarray*}
S_{i1} &=& \frac{S_{01}}{k_i+t}H_+ \label{5S1}, \quad 
 D_{i1}~ = \frac{D_{01}}{k_i-t}H_-. \label{5D1},\quad S_{i2}=D_{i2}=0
\\
H_+^{-1} &=& \frac{m_1}{k_1+t}+\frac{m_2}{k_2+t}, \quad
H_-^{-1} =  \frac{m_1}{k_1-t}+\frac{m_2}{k_2-t}. 
\end{eqnarray*}
We compute  bound $B_E$:
\begin{eqnarray}
B_E &=&\max_{t\in [0,k_1)} (Y_E - 2 t r), 
\label{we} \\
Y_E&=&\frac{1}{2}\left[H_+(1+r)^2+ H_-(1-r)^2\right]
\nonumber
\end{eqnarray}
Optimal value $t_{opt}$ of $t$ in region E is a root of the equation $\frac{d\, {(Y_E-2tr)}}{d\, t}=0$ which is a fourth-order equation of $t$. $\phi_{D_2E}$ is found by solving $\displaystyle \frac{d\, {(Y_E-2tr)}}{d\, t}|_{t=k_1}=0$ for $r$. After $t_{opt}(r)$ is founded and substituted into \eq{we}, one can compute the G-closure boundary using \eq{k-l}. We are not showing this calculation here. 

\paragraph{Summary: Regions D and E}
The regions are curved tetragons located in the domain of larger $m_1$. Region D is adjoined to B along the curve $\psi_{BD}$, \eq{boundBD} and  E is adjoined to C along the curve $\psi_{CE}$, \eq{bound-CE}.
These two regions are divided by the curve 
\begin{eqnarray}
 \phi_{D2E} &=&-\frac{m_1b-2m_2k_1 \tilde{k}+a\sqrt{m_1\left(m_1-1\right)b}}{2m_2k_1\tilde{k} }
 \label{psi-E}\\
& & a=\tilde{k}+\left(m_1+m_2 \right)k_1,  \nonumber\\ 
& &~b=a^2-(k_1+k_2)^2m_1-4m_2{k_1}^2 \label{pd2} \nonumber,
\end{eqnarray}
and are described as 
\begin{eqnarray*}
\Phi_{D} &:&~   \phi_{DB} < r, \quad \phi_{D2E}  \le  r\leq 1,~~ m_1 \leq 1-m_2, 
\\
\Phi_E &:& \max\{0, \phi_{CE}\}< r < \phi_{D2E}, \quad   m_1 \leq 1-m_2,
\end{eqnarray*}

\section{Optimal structures}\label{s:struct}

Here, we describe structures that realize the bounds. All of them are laminates of a rank, obtained by sequentially adding to the existing laminate a new one. The normal of laminates always coincide with one of mutually orthogonal axis $x_1$ and $x_2$. The fields in the materials is also codirected with these axes everywhere, they are denoted as $e_{ij}=[ \alpha_n, \beta_n]$ where the first index $i$ shows the location of the layer, second index $j$ shows the material, $\alpha_n$ and $\beta_n$ show the intensities of the field along the $x_1$ and $x_2$ directions in a laminate labeled $n$, respectively. The relations between parameters $\alpha_n, \beta_n$ and $s_n, d_n$ are given by  \eq{x2}. The plane of eigenvalues $(\alpha, \beta)$ is the rotated $45^\circ$ plain of $(s_{1}, d_{1})$. It such laminates, $d_{12}=d_{22}=0$ identically.
The field in the third material is always zero, which reads $ e_{i3}=[0,0]$.

We prove the optimality of the structure by a straight calculation of the fields $e_{ij}$ in materials in an optimal structure using the conditions of the rank-one connection - continuity of the tangent component of the fields in laminates and the sufficient optimality conditions found in Section \ref{s:bounds} .

\subsection{Intermediate Region B}\label{s:structB}

The multiscale laminates ($L(13,2,13)$-structure) that connect the fields satisfying the sufficient conditions shown in Figure \ref{f:str}, region B is attained by the following steps.
\begin{enumerate}
\item
$L(13)$ substructure is formed: materials $k_1$ and $k_3$ is laminated along $x_2$ direction with relative volume fraction of material $k_1$ equaling to $\mu_{11}$. The fields in the materials are called $e_{11}$ and $e_{13}$, and $e_{13} =0$. The average field in the substructure is called $e_{10}$ 

\item
$L(13,2)$ substructure is formed: The obtained $L(13)$ structure is laminated with material $k_2$  (field $e_{22}$) along $x_{1}$ direction with relative volume fraction of material $k_2$ equaling to $\mu_2$ to form a second rank laminate. The average field in the substructure is called $e_{20}$.

\item $L(13,2,13)$ substructure is formed: The obtained $L(13,2)$ structure is laminated along $x_2$ direction  with another laminate which is formed of material $k_1$  (field $e_{31}$) and $k_3$ (field $e_{33}=0$) with laminating direction parallel to $x_1$ and relative volume fraction of material $k_1$  equaling to $\mu_{31}$. The relative volume fraction of the second rank laminate is $\mu_4$. The average field in the substructure is called $e_{30}$. 

\end{enumerate}

The average fields in the described substructures are
\begin{eqnarray}
e_{10}&=&e_{11} \mu_{11} \label{eq2}\, \quad
e_{20}=e_{10} (1-\mu_2)+e_{22} \mu_{2}, \label{eq4}\quad
e_{30}=e_{31} \mu_{31} (1-\mu_4)+e_{20} \mu_4.  \label{eq6}
\end{eqnarray}
The laminate's volume fractions satisfy the geometric constraints:
\begin{eqnarray}
\mu_{11} (1-\mu_2)  \mu_4 + \mu_{31} (1-\mu_4) &=& m_1, \quad
\mu_4  \mu_2 = m_2 \label{ref4}
\end{eqnarray}
We compute the fields $e_{ij}$ and volume fractions $\mu_{ij}$ that satisfy both rank-one conditions and sufficient optimality conditions \eq{opt_B} found in Section \ref{s:t=k2}. Thus we show the realizability of the bound and find the applicability region of the bound.
 
Notice that material $k_1$  is always laminated with material $k_3$ with zero field;  using the rank-one connection - continuity of the tangential component of the fields in laminates, we see that the first entry of $e_{11}$ and the second entry  of $e_{31}$ are zeros; such fields satisfy sufficient condition \eq{det0}.
The sufficient optimality condition \eq{opt_B} requires that the the magnitude of $e$ is constant in $\Omega_1$,  therefore
\begin{eqnarray*}
e_{11} =[0, \quad  \beta], \quad 
e_{31}=[ \beta, \quad 0] 
\end{eqnarray*}
where $\beta $ is a constant. 
Optimality condition \eq{opt_B} also requires that field in Material $k_2$  is proportional to identity matrix:
\begin{eqnarray}
e_{22} =[\alpha_1, \quad \beta_1 ], \quad \alpha_1=\beta_1 
\label{e2}
\end{eqnarray}

The remaining group of continuity conditions concerns the average field $e_{i0}$ in the substructures. Based on the rank one connection condition we have the following: 
\begin{eqnarray}
e_{10}[2] &=&e_{22}[2] ~ \to ~\beta \mu_{11} = \beta_1 \\
e_{20}[1] &=&e_{30}[1]~\to ~\alpha_1  \mu_2 = \beta \mu_{31} 
\end{eqnarray}
The average field of the mixture equaling to external field leads us to:  $e_{30} = [ 1, \quad r]$, hence:
\begin{eqnarray}
1=\alpha_{1}  \mu_2 (1-\mu_5)+ \beta \mu_{31} \mu_5  \nonumber \\
 (\beta  \mu_{11} (1-\mu_2)+ \beta_1 \mu_2) \mu_4 = r   \label{volume2} 
\end{eqnarray}

\paragraph{Constraints and parameters of an optimal laminate}

Solving \eq{ref4}, \eq{e2}-\eq{volume2}, we obtain:
\begin{eqnarray*}
\beta &=& \frac {(1+r)-2 \sqrt{m_2r}}{m_1}, \quad
\alpha_1 = \beta_1 = \sqrt{\frac{r}{m_2}} \\
\mu_{11} &=& \frac {rm_1}{(1+r)\sqrt{rm_2}-2r\,m_2}, \quad \mu_4 =\sqrt{rm_2} \\
\mu_{31} &=& \frac {m_1}{(1+r)-2\sqrt{rm_2}}, \quad
\mu_2 = \sqrt{\frac{m_2}{r}}
\end{eqnarray*}
A straight calculation confirms that the fields coincide with the fields computed for the bounds. 
\paragraph{Region of applicability}
Requiring all volume fractions fall into the interval (0,1) and also enforcing $0<r\le 1$, we have a system of inequalities of $r$:
\begin{eqnarray*}
0 < \frac {rm_1}{(1+r)\sqrt{rm_2}-2r\,m_2} <1, \nonumber \\
0 < \frac {m_1}{(1+r)-2\sqrt{r\,m_2}} <1, \nonumber \\
0 < \sqrt{\frac {m_2}{r}} <1, \quad
0< \sqrt{rm_2} <1 \label{app_B}
\end{eqnarray*}
Solve the above inequalities and we found that:
\begin{eqnarray*}
 \left\{ \begin{array}{ll} \ m_2< r < 1 \quad &\mbox{if }~ m_1 < 2\left(\sqrt{m_2}-m_2\right), \\  
\displaystyle m_2 <r <\frac{1 -a_5 - \sqrt{1- 2 a_5}}{a_5} ,  & \mbox{otherwise }. \end{array} \right.
\end{eqnarray*}
which is consistent with the region of applicability of the bound in region B.
\paragraph{Region C} Note that if $r=m_2$, then $\mu_2=1$, which implies that the inner layer of composite disappears or the composite degenerate into $T$ structure - second-rank laminate $L(13,2)$ Figure \ref{f:str}, region C, that matches the bound in region C.  This structure plays the same role as laminate in two-phase problem. There, an optimal structure degenerates into laminate if $ r$ is small enough.

\subsection{Region A}\label{s:structA}

\paragraph{Region $A_1$: L(13,2,13,2)-structure:}
Laminates  L(13,2,13,2) whose field inside each phase match  the sufficient conditions shown in Figure \ref{f:str}  region A are obtained by adding a layer of material $k_2$   (field $e_{42}$) along direction $x_1$ to the  laminate  $L(13,2,13)$ described in previous section,  with the relative volume fraction of $L(13,2,13)$ defined as $\mu_5$.  The average field $e_0$ in the structure is again $e_0=[1, r]$.
The average fields in the substructures are as in \eq{eq6} and additionally we have $ e_{40}=e_{30} \mu_5 + e_{42} (1-\mu_5) $.
Continuity conditions for average field $e_{i0}$ in the substructures are
\begin{eqnarray*}
e_{10}[2] =e_{22}[2],  \quad
e_{20}[1] = e_{31}[1] \mu_{31}, \quad
e_{30}[2]=e_{42}[2] 
\end{eqnarray*}
Average field $e_{40}=[1,r]$ leads to:
\begin{equation}
1=\alpha_2 (1-\mu_5)+ \beta \mu_{31} \mu_5 \label{volume1a}
\end{equation}
The volume fractions (relative and absolute) are related as:
\begin{eqnarray}
m_1 =\mu_{11} (1-\mu_2) \mu_4 \mu_5+\mu_{31} (1-\mu_4) \mu_5, \label{eq9}\quad
m_2 = \mu_2 \mu_4 \mu_5+(1-\mu_5)\label{eq10}
\end{eqnarray}
The optimality conditions \eq{Si1_A}, \eq{opt_A} yield to relations
\begin{eqnarray*}
& & e_{11} =[0, ~  \beta], \quad
e_{31}=[ \beta, ~ 0], \quad 
e_{22} =[\alpha_1, ~ \beta_1 ], \quad e_{42}=[\alpha_2, ~ r],
\\
& &\alpha_1 + \beta_1 =\alpha_2 + r 
\quad
k_1 \beta ={k_2}(\alpha_1 + \beta_1 )
\end{eqnarray*}
where  $\alpha_1, \alpha_2, \beta, \beta_1$ are some constants. 
The established relations allow for solving for the unknown parameters of the structure - the constants $\alpha_1, \alpha_2, \beta, \beta_1$ and $\mu_{ij}$. 
\begin{remark}Here, the number of unknowns is bigger than the number of constants. To handle the uncertainty, we additionally assume that $e_{22}$ is proportional to identity matrix, i.e. $e_{22}[1]=e_{22}[2]$ or 
\begin{eqnarray}
\alpha_1=\beta_1 \label{eq8},
\end{eqnarray}
which significantly simplifies the calculations but might restrict the domain of applicability.    

 L(13,2,13,2,2)-structure is optimal in the isotropic case ($r=1$) and  the field inside material $k_2$  in the core part is proportional to identity matrix (Cherkaev, 2009). The suggested here L(13,2,13,2)-structure is a degeneration of the L(13,2,13,2,2)-structure, so we keep the assumption on the field inside core material $k_2$. 
\end{remark}

\paragraph{Calculation of the constants}
Solving equations (\ref{volume1a})-(\ref{eq8}), we find the volume fractions and fields inside each material:
\begin{eqnarray*}
\mu_{11}&=&\frac{k_1}{2k_2}, \quad
\mu_{31}=\frac{a_2\,k_1^2}{r\, k_2 a_1},\quad
\mu_2 = \frac{2a_2 k_1}{r\, a_1},\\
\mu_4 &=& \frac{2 \tilde{k}}{k_1 (r+1)},\quad
\mu_5 = \frac{r(1+r) a_1}{(2 \tilde{k}- k_1(1+r))^2}.\\
e_{11} &=& \frac{k_2 (r+1)}{\tilde{k}}\left[ 0 , \, 1  \right], \quad
e_{31} =  \frac{k_2 (r+1)}{\tilde{k}} \left[1, \, 0 \right], \\
e_{22} &=&\frac{k_1 (r+1)}{2 \tilde{k}} \left[ 1,\, 1 \right], \quad
e_{42} =\frac{k_1 (r+1)}{\tilde{k}}[1,\,0] + r [-1, 1] 
\end{eqnarray*}
where:
\begin{eqnarray*}
 a_1 &=& -5r m_2{k_1}^2 - 4r \,k_1m_1k_2+4r \tilde{k}^2 +(1+r-m_2){k_1}^2 
\\
a_2 & =& k_1{m_2}(m_2-1) +m_1 k_2(m_2+r)
\end{eqnarray*}
Notice that  fractions $\mu_2$ and $\mu_{31}$ vanish simultaneously. Such degeneration brings the  $ L(13,2,13,2)$ structure  into $ L(13,2)$-structure\\

\paragraph{Region of applicability}
All the volume fractions have to fall in the the interval (0,1), therefore we have the following inequalities:
\begin{eqnarray*}
0 < \frac{(m_2 \tilde{k}-k_1m_2+r m_1k_2){k_1}^2}{r\,k_2 a_2} <1 \label{eq11}\\
0 < \frac{2(m_2 \tilde{k}-k_1m_2+r m_1k_2) k_1}{r\,a_2} < 1 \nonumber \\
0 < \frac{2  \tilde{k}}{k_1(r+1)} < 1, \quad
0 < \frac{r(1+r)a_2}{(-r\,k_1-k_1+2 \tilde{k})^2} < 1 \label{eq12}
\end{eqnarray*}
The above system of inequalities have solutions:
\begin{eqnarray}
 \psi_{A1A2} < r < \psi_{A1B},~~ \label{boundA1A2}
\end{eqnarray}
where
$$
 \psi_{A1A2}=\frac{(k_1- \tilde{k})m_2}{m_1k_2} , \quad \psi_{A1B}=\frac {m_2k_1^2}{2  \tilde{k}^2-m_2k_1^2+2 \tilde{k}\sqrt{ \tilde{k}^2 -m_2k_1^2}}
$$
only if $k_i, ~m_i, i=1,2$ satisfy:
$$
m_1 <\frac {k_1(1-m_2)}{2k_2}.
$$
At the boundary of the applicability domain where  $r=\psi_{A1B}$ we compute $\mu_5=1$ and conclude that the structure degenerates into $L(13,2,13)$-laminate. At the other boundary, when $ r=\psi_{A1A2}$ we have $\mu_2=0, \mu_{31}=0$ which means that the composite degenerates into $L(13,2)$-structure.

\paragraph{Region A2: L(123,2)-structure} In the region A2, the optimal structure is second-rank laminate $L(123,2)$, see figure \ref{f:str}region A2. The structure contains the following fields: In the inner layer,
$ e_1 = (\alpha_1,0)$ in $\Omega_1$, $e_{12}  = (\alpha_2,0)$ in $\Omega_{21}$, and $ e_3 = (0,0)$ in $\Omega_3$, and the in outer layer, $e_{22}  = (\alpha_{22},\, \beta_{22})$ in $\Omega_{22}$. Here $\Omega_{21}$ and $\Omega_{22}$ are the subdivisions of $\Omega_{2}$, $\Omega_{2}=\Omega_{21}\cup \Omega_{22}$.

The optimality condition requires that $ \alpha_2= \alpha_{22}+ \beta_{22} $ and the compatibility requires that $ k_1 \alpha_1=k_2 \alpha_2. $. The computation of the constants is similar to the previous case. They are:
\begin{eqnarray*}
\alpha_1= \frac{k_2(1+r)}{\tilde{k}}\quad
\alpha_2= \frac{k_1(1+r)}{\tilde{k}}
\\
\alpha_{22}=r \quad \beta_{22}=\frac{k_1 (1+r) }{\tilde{k}} -r
\end{eqnarray*}

The region is bounded by the lines, see Figure \ref{f:allregions}.
$$r=0. ~~ r=1, ~~ m_1=0, ~~r=\psi_{A1A2}, ~~r=\psi_{AC} 
$$
When $r\to 0$, the optimal  L(123,2)-structure degenerates into laminate.

\subsection{Region D}\label{s:struct-d}

A part D1 of region D is attainable by laminates L(13,2,13,1,1), as it is shown in (Albin et al. (2007)) by the method similar to the presented above. The most anisotropic structure of this type is L(13,2,1) with the parameters:
\begin{eqnarray}
p=\frac{k_1(1-m_2-m_1)}{m_1k_2},\quad
e_{11}=\left[ 0, \, \frac{r (k_1+k_2) (k_1+k_2- \tilde{k}-m_1k_1)}{m_2k_2k_1}\right], \nonumber \\
e_2=\left[ \frac{r (k_1+k_2- \tilde{k}-m_1k_1)}{m_2k_2}, \, \frac{r (k_1+k_2- \tilde{k}-m_1k_1)}{m_2k_2} \right],\nonumber \\
e_{31}=\left[r, \, \frac{r(1-m_1)(k_1+k_2)^2}{m_2k_1k_2}-\frac{r (k_1+2k_2)}{k_2} \right].\nonumber
\end{eqnarray}
and $pm_1$ is the volume fraction of material $k_1$  that is rank-one connected with material $k_3$.
L(13,2,1) is optimal when
\begin{eqnarray}
 r &=& \psi_{D1D2},  \nonumber \\ 
\psi_{D1D2}&=& \frac{m_2k_1k_2}{(k_1+k_2) m_1((k_1+k_2)(1-m_1)-3m_2k_1)+ m_2k_1(2k_1+k_2-2 m_2k_1)}
\label{psi-d}
\end{eqnarray}
and region $D1$ is described as
\begin{equation}
 m_1\geq \frac{k_1(1-m_2)}{k_1+k_2}, \quad \psi_{D1D2} \leq r\leq 1,~~ r \ge \psi_{BD}.
 \label{dd1}
\end{equation}
In Region D2 where $r<\psi_{D1D2}$, we did not find optimal structures. We presume that the translation bound is rough in that region. Our guess is to preserve structures L(13,2,1) as the best ones (see discussion in the next Section).

\subsection{Attainability of bounds in D and E regions}\label{s:structE}

The bound \eq{we} in region E is not attainable. Indeed, it suggests that the local fields are constant in each material. These constant fields, however, cannot be joined together with a structure. Indeed, an arbitrary mixture of fields $ E_1$ and $E_2$ cannot be in connection with $E_3=0$. Indeed,  the determinant of the difference of two connected fields must be zero. But one can check by direct calculation that determinant of any convex combination of $ c\,E_1+ (1-c)E_2$ from \eq{5S1}, \eq{5D1} is not zero. The bound cannot be exact: the field in $\Omega_3$ can neighbor neither fields in $\Omega_1$ and $\Omega_2$ nor their mixture.

\begin{remark}
An exact bound would require more restrictive constraint  than \eq{Nesiineq}  $\det(e(x))\geq 0$ used here. These new  constraints should depend on volume fractions of materials in the composite or on the mean field in it. At present, the needed constraints are not established and the obtained bound \eq{we} is rough (not attainable) although very close, see Section \ref{s:structE}.
\end{remark}

The part of region D (region D2) close to E  is probably not attainable as well. As we described earlier, a larger part D1 of region D is attainable by structures are L(13, 2, 13, 1,1), Figure \ref{f:lm2}, (Albin et al. (2007)). In D1, values of  $m_1$ and $r$ are larger than the corresponding thresholds. Region D1 adjoins region B along curve $\psi_{BD}$, see figure \ref{f:allregions}.   The most anisotropic structures of this sort are L(13, 2, 1) (Albin et al. (2007)), they correspond to minimal values $r=\phi_{D_1{D_2}}$ computed in (Albin et al. (2007)).
\begin{eqnarray}
\phi_{D1D2}= \frac{m_2k_1k_2}{(k_1+k_2) m_1((k_1+k_2)(1-m_1)-3m_2k_1)+ m_2k_1(2k_1+k_2-2 m_2k_1)} \label{eq_BD}
\end{eqnarray}
 We have not found optimal structures in the complementary region $D2$ that is between regions D1 and $E$ and conjecture that the bound probably is not exact there. In this region $t_{opt}=k_1$ but the bound is not attainable. Probably, a better bound is needed for this region and for region E.

\paragraph{Arguments for presumptive structures} We try to guess the best structures in D2- and E-regions.
The bound is not exact, and we would not attempt to use the method that was exploited for the other regions: there is no chance to find the structure that exactly realizes the bound. Instead, we conjecture the type of laminate structures based on their asymptotic behavior, adjust an inner structural parameter to the external field, and compare the result with the bound, finding the gap between them.

We notice that at the boundary of neighboring region $D1$, optimal structures are of the type $L(13,2,13,1)$  degenerate on the boundary into the structures $ L(13,2,1)$, Figure \ref{f:allregions}. 
In another neighboring region C, optimal structures $L(13,2)$ can be viewed as  the result of degeneration of  $ L(13,2,1)$, when the volume fraction of the exterior layer goes to zero.  

Also, laminates $L(12)$ are optimal two-material ($m_3\to 0$) structures that correspond to anisotropic loading $r<r_{tr}<1$, (Lurie and Cherkaev, 1982), these laminates are a degeneration of laminates $L(13,2,1)$ as $m_3 \to 0$.
Finally, the optimal structures for the limits when $r\to 0$ where conjuncted in (Cherkaev and Gibiansky, 1988) to be $ L(13,2,1)$, because this structure, being different from simple laminates, has a conductivity in $x_1$ direction equal to harmonic mean  while the conductivity in orthogonal direction is finite (smaller than the arithmetic mean). 

Based on these observation, we presume that structures $ L(13,2,1)$ with the proper distribution of $k_1$ between the layers, stay optimal in the whole region $D2 \cup E$.  In those structures, the fields in the second and third materials are constant everywhere, as the bound predicts. However,  the field in the first material  takes two different values in different layers; this contradicts the assumption of the bound but makes the structure compatible: The field $e_{11}$ in the inner layer is rank-one connected with $E_3$.

\paragraph{Numerical results}

\begin{figure}[htbp]
\begin{center}
\includegraphics[scale=.55]{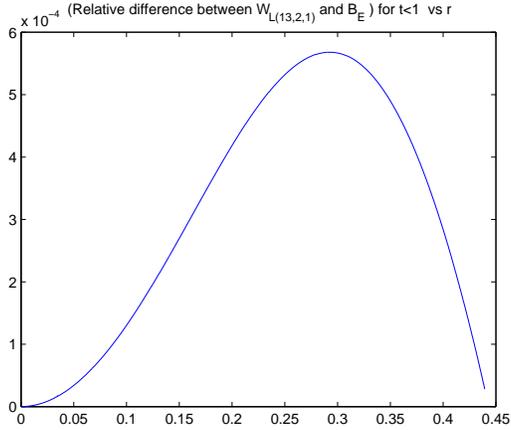}
\caption{{$\delta W_{rel}$ \bf relative gap between the energy of the bounds and guessed structure in region E.}}
\label{f3}
\end{center}
\end{figure}

The numerical experiments are performed to see how well the suggested bounds approximate an optimal bound. In all the numerical experiments, conductivities $k_i$ of each material is fixed. So is the volume fraction $m_i$ of each material. And relative difference between the bound $B_E$ and energy $W_L(13,2,1)$ of L(13,2,1) structures (Figure \ref{f:str}, region  E) as follows
$$
\delta W_{rel}=  \left(  \frac{\min_{\alpha\in [0, 1]}W_{L(13,2,1)}(\alpha)  -B_E}{B_{E}}\right)
$$
are calculated for $r\in [0,r_0]$, where $r_0$ is the threshold value where the $t_{opt}$ in case E becomes $k_1$.
Energy $W_{L(13,2,1)}$ depends on one parameter $\alpha$ - the relative amount of material one used in the inner layer.
We choose value $\alpha_{opt}$ of $\alpha$ to minimize the energy stored in the structure. $\alpha_{opt}$ changes with respect to anisotropy level $r$ of external field hence $W_{L(13,2,1)}$. 
The results of one numerical experiment are showed in Figure \ref{f3} and the parameters are: $m_1=0.2,  m_2=0.5,    m_3=0.3, k_1=1, k_2=3$.  As we can see, the relative differences are rather small, are in order of $10^{-4}$ and even in order of $10^{-7}$ as $r$ is very close to value of 0. This is true for all fixed $m_1$ values which fall inside region $E$ in $r-m_1$ plane and $\delta W_{rel}$ also changes in the same way with respect to $r$ for each fixed $m_1$.  

\paragraph{Acknowledgment} The research is supported by the grant from  NSF: DMS-0707974.

\end{document}